\documentclass{aa}
\usepackage{graphicx}
\usepackage{xcolor}
\usepackage{txfonts}
\usepackage[colorlinks=true,citecolor=cyan]{hyperref}
\def\Edit{\textcolor{black}}
\def\Editm{\textcolor{black}}

\begin{document} 

   \title{A puzzling non-detection of [\ion{O}{iii}] and [\ion{C}{ii}] from a $z \approx 7.7$ galaxy observed with ALMA}

    \titlerunning{A puzzling non-detection of [\ion{O}{iii}] and [\ion{C}{ii}] from a $z \approx 7.7$ galaxy}

   \author{C. Binggeli \inst{1} \and A. K. Inoue \inst{2,3,4} \and T. Hashimoto \inst{3,5,4} \and M. C. Toribio \inst{6} \and E. Zackrisson \inst{1} \and S. Ramstedt \inst{7} \and K. Mawatari\inst{8,4} \and Y. Harikane\inst{9,10} \and H. Matsuo \inst{10,11} \and T. Okamoto \inst{12} \and K. Ota \inst{13} \and I. Shimizu \inst{10,14} \and Y. Tamura \inst{15} \and Y. Taniguchi \inst{16} \and H. Umehata \inst{17,18}}

   \institute{Observational Astrophysics, Department of Physics and Astronomy, Uppsala University, Box 516, SE-751 20 Uppsala, Sweden\\
            \email{christian.binggeli@physics.uu.se}
            \and
            Department of Physics, School of Advanced Science and Engineering, Waseda University, 3-4-1 Okubo, Shinjuku, Tokyo 169-8555, Japan
            \and
            Waseda Research Institute for Science and Engineering, 3-4-1 Okubo, Shinjuku, Tokyo 169-8555, Japan
            \and
            Department of Environmental Science and Technology, Faculty of Design Technology, Osaka Sangyo University, 3-1-1, Nakagaito, Daito, Osaka, 574-8530, Japan
            \and
            Tomonaga Center for the History of the Universe (TCHoU), Faculty of Pure and Applied Sciences, University of Tsukuba, Tsukuba, Ibaraki 305-8571, Japan
            \and
            Department of Space, Earth and Environment, Chalmers University of Technology Onsala Space Observatory, SE-439 92 Onsala, Sweden
            \and 
            Theoretical Astrophysics, Department of Physics and Astronomy, Uppsala University, Box 516, SE-751 20 Uppsala, Sweden
            \and
            Institute for Cosmic Ray Research, The University of Tokyo, Kashiwa, Chiba 277-8582, Japan
            \and
            Department of Physics and Astronomy, University College London, Gower Street, London WC1E 6BT, UK
            \and
            National Astronomical Observatory of Japan, 2-21-1 Osawa, Mitaka, Tokyo 181-8588, Japan
            \and
            Department of Astronomical Science, The Graduate University for Advanced Studies, 2-21-1 Osawa, Mitaka, Tokyo 181-8588, Japan
            \and
            Faculty of Science, Hokkaido University, N10 W8, Kitaku, Sapporo, Hokkaido 060-0810, Japan
            \and
            Kyoto University Research Administration Office, Yoshida-Honmachi, Sakyo-ku, Kyoto 606-8501, Japan
            \and
            Shikoku Gakuin University, 3-2-1 Bunkyocho, Zentsuji, Kagawa 765-0013, Japan
            \and
            Division of Particle and Astrophysical Science, Graduate School of Science, Nagoya University, Aichi 464-8602, Japan
            \and
            The Open University of Japan, 2-11 Wakaba, Mihama-ku, Chiba 261-8566, Japan
            \and
            RIKEN Cluster for Pioneering Research, 2-1 Hirosawa, Wako-shi, Saitama 351-0198, Japan
            \and
            Institute of Astronomy, School of Science, The University of Tokyo, 2-21-1 Osawa, Mitaka, Tokyo 181-0015, Japan
            }
            
\authorrunning{C. Binggeli et al.}

   \date{Received September XX, XXXX; accepted March XX, XXXX}

 
  \abstract
   {Characterizing the galaxy population in the early Universe holds the key to understanding the evolution of these objects and the role they played in cosmic reionization. However, the number of observations at the very highest redshifts are to date, few.}
   {In order to shed light on the properties of galaxies in the high-redshift Universe and their interstellar media, we observe the Lyman-$\alpha$ emitting galaxy \object{z7\_GSD\_3811} at $z=7.664$ with band 6 and 8 at the Atacama Large Millimeter/submillimeter Array (ALMA).}
   {We target the far-infrared [\ion{O}{iii}] 88 $\mu \mathrm{m}$, [\ion{C}{ii}] 158 $\mu \mathrm{m}$ emission lines and dust continuum in the star-forming galaxy z7\_GSD\_3811 with ALMA. We combine these measurements with earlier observations in the rest-frame ultraviolet (UV) in order to characterize the object, and compare results to those of earlier studies observing [\ion{O}{iii}] and [\ion{C}{ii}] emission in high-redshift galaxies.}
   {The [\ion{O}{iii}] 88 $\mu \mathrm{m}$ and [\ion{C}{ii}] 158 $\mu \mathrm{m}$ emission lines are undetected at the position of z7\_GSD\_3811, with $3\sigma$ upper limits of \Edit{$1.6 \ \times \ 10^{8} \ \mathrm{L_{\odot}}$} and $4.0 \ \times \ 10^{7} \ \mathrm{L_{\odot}}$, respectively. We do not detect any dust continuum in band 6 nor band 8. The measured rms in the band 8 and band 6 continuum is 26 and 9.9 $\mathrm{\mu Jy \ beam}^{-1}$, respectively. Similar to several other high-redshift galaxies, z7\_GSD\_3811 exhibits low [\ion{C}{ii}] emission for its star formation rate compared to local galaxies. Furthermore, our upper limit on the [\ion{O}{iii}] line luminosity is lower than all the previously observed [\ion{O}{iii}] lines in high-redshift galaxies with similar ultraviolet luminosities. Our ALMA band 6 and 8 dust continuum observations imply that z7\_GSD\_3811 likely has a low dust content, and our non-detections of the [\ion{O}{iii}] and [\ion{C}{ii}] lines could indicate that z7\_GSD\_3811 has a low metallicity ($Z \lesssim 0.1 \ \mathrm{Z_{\odot}}$).}
{}

   \maketitle
%

\section{Introduction}
\label{sec:introduction}

\Edit{In order to provide a complete picture of the cosmic reionization and the evolution of galaxies across cosmic time, we need to understand the nature of the galaxy population present during the first billion years after the Big Bang.} The properties of these objects and their interstellar media (ISM) are, however, still poorly understood, and characterizing physical properties of the galaxy population at the very highest redshift remains a challenging task. This is not least due to the lack of observational facilities that can provide us with high-quality spectra in the rest-frame optical and ultraviolet (UV), but also due to the limited number of currently observable features that allow for reliable spectroscopic confirmation of possible high-redshift candidates. While the challenge of obtaining high-quality spectra in the rest-frame optical and UV may have to wait until the launch of the James Webb Space telescope (JWST), the Atacama Large Millimeter/submillimeter Array (ALMA) has provided us with an excellent way to probe these galaxies and their ISM through far-infrared (FIR) emission lines and FIR dust emission, since these features fall into ALMA wavelengths at high redshifts. 

In recent years, many ALMA studies of high-redshift galaxies have been aimed at observing two such FIR emission lines: the [\ion{C}{ii}] 158 $\mu \mathrm{m}$ line \citep[see e.g.][for a summary of such observations]{carniani_kiloparsec-scale_2018, matthee_resolved_2019, harikane_large_2019}, and the [\ion{O}{iii}] 88 $\mu \mathrm{m}$ line \citep{inoue_detection_2016,laporte_dust_2017,carniarni_extended_2017,hashimoto_onset_2018,tamura_detection_2019,hashimoto_big_2019,harikane_large_2019}. 
While the [\ion{C}{ii}] line arises predominantly in photo-dissociation regions (PDRs), where hydrogen is neutral, the [\ion{O}{iii}] line originates in \ion{H}{ii}-regions \citep[e.g.][]{abel_region_2005}. These lines can thus be used to examine a range of properties, such as the ionization state and chemical properties at high redshifts \citep[e.g.][]{vallini_CII_2015,vallini_molecular_2017,harikane_silverrush_2018,katz_probing_2019,harikane_large_2019}. For example, high [\ion{O}{iii}]-to-[\ion{C}{ii}] ratios may be an indication of a low neutral hydrogen content, which is associated with high Lyman continuum escape fractions \citep[e.g.][]{inoue_detection_2016}, and thus could help us understand the reionization of the Universe.

\Edit{A number of studies targeting the [\ion{C}{ii}] 158 $\mu \mathrm{m}$ line in high-redshift star-forming galaxies have reported surprisingly weak lines compared to local objects with similar star formation rates \citep[SFRs; e.g.][]{Ota_ALMA_2014,Schaerer_constraints_2015,maiolino_assembly_2015}. In their analysis of a compilation of galaxies at $z\sim 5\text{ -- }9$, \citet{harikane_large_2019} found that these exhibit a deficit in [\ion{C}{ii}] with respect to the local [\ion{C}{ii}]-SFR relation. They also found that this deficit increases with stronger $\mathrm{Ly\alpha}$ emission, which is consistent with the results of \citet{carniani_kiloparsec-scale_2018, harikane_silverrush_2018}. Several mechanisms have been proposed in order to explain the origin of weak [\ion{C}{ii}] emission in high-redshift galaxies, such as low metallicity, increased ionization parameters, lower PDR covering fraction or increased PDR densities in combination with a lower C/O abundance ratio or cosmic microwave background (CMB) effects \citep[e.g.][]{vallini_CII_2015,harikane_silverrush_2018,harikane_large_2019}. It has also been suggested that the possible weakness of [\ion{C}{ii}] emission in high-redshift galaxies may be a result of a selection bias due to selecting in the UV \citep{katz_probing_2019}.} \Edit{On the other hand, when analysing a compilation of $z \approx 6 \text{ -- } 7$ galaxies similar to the one used by \citet{harikane_large_2019}, \citet{matthee_resolved_2019} do not find that high-redshift galaxies deviate from the local [\ion{C}{ii}]-SFR relation at $\mathrm{SFR} \gtrsim 30 \ \mathrm{M_{\odot}\ yr^{-1}}$. They do, however, find a possible deviation at lower SFR. \citet{schaerer_ALPINE_2020} also find good agreement between $z\sim 4 \text{ -- } 6$ star-forming galaxies from the APLINE-ALMA survey \citep{lefevre_ALPINE_2019,bethermin_ALPINE_2020,faisst_ALPINE-ALMA_2020} and local galaxies. When combined with earlier measurements at $z>6$, they find only a slightly steeper [\ion{C}{ii}]-SFR relation than the local one. Furthermore, in contrast to \citet{harikane_large_2019}, \citet{schaerer_ALPINE_2020} do not find a strong dependence of $L_{\mathrm{[\ion{C}{ii}]}}/\mathrm{SFR}$ on the $\mathrm{Ly\alpha}$ equivalent width.}

\Edit{The [\ion{O}{iii}] 88 $\mu \mathrm{m}$ emission line has been observed} to be brighter than the [\ion{C}{ii}] 158 $\mu \mathrm{m}$ line in low-metallicity dwarf galaxies \citep[e.g.][]{cormier_herschel_2015} and is expected to also be so in the high-redshift Universe where metallicities generally should be lower \citep{inoue_ALMA_2014}. Indeed, as pointed out by \citet{hashimoto_big_2019}, the hunt for [\ion{O}{iii}] 88 $\mu \mathrm{m}$ emission in galaxies $z>6$ has thus far been extremely successful, with successful detections all the way up to $z\sim 9$ \citep{hashimoto_onset_2018}. Recent results also seem to indicate that [\ion{O}{iii}] emission in high-redshift galaxies is in better agreement with local galaxies with similar SFRs compared to the [\ion{C}{ii}] line, suggesting a relatively tight correlation between SFR and [\ion{O}{iii}] emission even at $z>6$ \citep{harikane_large_2019}. The number of studies aimed at observing the [\ion{O}{iii}] line at high redshifts is, however, to date, significantly smaller than those targeting the [\ion{C}{ii}] line.

In this study, we present ALMA observations targeting [\ion{C}{ii}] 158 $\mu \mathrm{m}$, [\ion{O}{iii}] 88 $\mu \mathrm{m}$ and dust continuum emission in the $z=7.664$ $\mathrm{Ly\alpha}$-emitting galaxy z7\_GSD\_3811 \citep{song_keck_2016}. The paper is organized as follows: In Sect.~\ref{sec:target} we briefly describe the target object. In Sect.~\ref{sec:ALMA_b6_b8_obs}, we describe our ALMA band 6 and band 8 observations of [\ion{C}{ii}] and [\ion{O}{iii}] emission, respectively. In Sect.~\ref{sec:results} we present our results from ALMA observational constraints from combining these with rest-frame UV data. In Sect.~\ref{sec:SED_fitting}, we present results from spectral energy distribution (SED) fitting by combining earlier data with our ALMA data. Finally, in Sect.~\ref{sec:Disc} we discuss and summarize our findings. Throughout the paper we assume the following cosmological parameters: $H_0 = 70 \ \mathrm{km \ s^{-1} \ Mpc^{-1}}$, $\Omega_m = 0.3$, \Edit{$\Omega_{\Lambda}=0.7$}. We use the AB magnitude system \citep{oke_reference_1983}. All quoted uncertainties represent 68\% confidence intervals.

\section{Target}
\label{sec:target}
We present ALMA observations of the galaxy z7\_GSD\_3811 \citep{finkelstein_evolution_2015,song_keck_2016}, a bright ($\mathrm{F160W}=25.8 \text{ mag}$) Lyman break galaxy (LBG) in the Great Observatories Origins Deep Survey South \citep[GOODS-S;][]{giavalisco_great_2004} field. The object has been observed with the \textit{Hubble Space Telescope} (HST) and HAWK-I at the Very Large Telescope (VLT) within the Cosmic Assembly Near-infrared Deep Extragalactic Legacy Survey \citep[CANDELS;][]{grogin_candels_2011, koekemoer_candels_2011}. Spectroscopic follow-up observations using Keck/MOSFIRE have detected a $\mathrm{Ly\alpha}$ line at $z=7.6637 \pm 0.0011$ with a rest-frame equivalent width EW($\mathrm{Ly\alpha}$)$\  = 15.6^{+5.9}_{-3.6}$ \citep{song_keck_2016}, making it one of the highest redshift objects to be spectroscopically confirmed. The detected $\mathrm{Ly\alpha}$ line places z7\_GSD\_3811 at a redshift where the [\ion{O}{iii}] 88 $\mu \mathrm{m}$ and [\ion{C}{ii}] 158 $\mu \mathrm{m}$ emission lines fall within ALMA band 8 and 6, respectively. Furthermore, there are several earlier studies of high-redshift galaxies with similar UV luminosities \citep[$M_{\mathrm{UV}}-21.22_{-0.10}^{+0.06}$,][]{song_keck_2016} that have led to detections in [\ion{C}{ii}] \citep{pentericci_tracing_2016,carniani_kiloparsec-scale_2018} and [\ion{O}{iii}] \citep{inoue_detection_2016,tamura_detection_2019}. Thus, z7\_GSD\_3811 should be a suitable target for expanding the currently limited number of [\ion{O}{iii}]-detected objects, but also for examining the ISM in the high-redshift Universe through a combination of [\ion{O}{iii}], [\ion{C}{ii}] and dust continuum emission.

\section{ALMA observations and imaging}
\label{sec:ALMA_b6_b8_obs}
Our ALMA observations of z7\_GSD\_3811 were performed during April to September 2018 as part of ALMA cycle 5 (Project ID: 2017.1.00190.S, PI: A. Inoue). Observations were performed in ALMA bands 6 and 8 targeting the [\ion{C}{ii}] 158 $\mu \mathrm{m}$ and [\ion{O}{iii}] 88 $\mu \mathrm{m}$ emission lines along with dust continuum around 152 $\mu \mathrm{m}$ and 87 $\mu \mathrm{m}$ rest-frame. Four spectral windows (SPWs) centered at approximately 219.4, 221.1, 234.0 and 236.0 GHz, each with a bandwidth of 1875 MHz and a resolution of 3.9 MHz were used in band 6. For the band 8 observations, four SPWs were placed at approximately 391.6, 393.3, 403.7 and 405.5 GHz, each with a bandwidth of 1875 MHz and a resolution of 7.8 MHz. In band 6, this corresponds to a velocity bandwidth of $\sim 2400 \text{ -- } 2600 \ \mathrm{km\ s^{-1}}$ per SPW, and a velocity resolution of $\sim 5 \ \mathrm{km\ s^{-1}}$. In band 8, the velocity bandwidth is $\sim 1400 \ \mathrm{km\ s^{-1}}$ per SPW, and the velocity resolution is $\sim 6 \ \mathrm{km\ s^{-1}}$. In both bands, one of the SPWs was centered around the position of the expected emission line of interest ([\ion{C}{ii}] 158 $\mu \mathrm{m}$ at 219.37 GHz in band 6, and [\ion{O}{iii}] 88 $\mu \mathrm{m}$ at 391.63 GHz in band 8) given the $\mathrm{Ly\alpha}$ redshift, while a spare SPW was placed at higher frequency in case of a large blueshift relative to the $\mathrm{Ly\alpha}$ emission line. The two remaining SPWs in each band were used in order to get observations of the dust continuum around 152 $\mu \mathrm{m}$ and 87 $\mu \mathrm{m}$ in band 6 and 8 respectively. z7\_GSD\_3811 was observed in band 6 during April and September of 2018, with a median precipitable water vapor (PVW) of $1.3 \text{ -- } 1.8$ mm, and in band 8 during August and September the same year, with a median PVW of $0.3 \text{ -- } 0.7$ mm.  A summary of the observations is presented in table~\ref{table:ALMA_observations}.  

\Edit{For} the band 6 observations, \Edit{the quasar (QSO)} J0522-3627 was used for bandpass and flux calibration for the executions performed in April 2018, while \Edit{the QSO} J0423-0120 was used for the execution performed in September 2018. In band 8, J0522-3627 was used for bandpass and flux calibration for all executions. In both bands, \Edit{the QSO} J0348-2749 was used for phase calibration for all the executions. We also use archival band 6 data from cycle 3 (Project ID: 2015.1.00821.S, PI: S. Finkelstein) observed in March and May 2016. The setup of the SPWs in these observations is similar to the setup used in the cycle 5 observations, with a total of four SPWs with bandwidths of 2000 MHz centered at similar frequencies (see table~\ref{table:ALMA_observations}). However, these observations were performed with a lower spectral resolution (15.6 MHz, corresponding to $\sim 20 \ \mathrm{km\ s^{-1}}$). For the execution performed in March 2016, \Edit{the QSO} J0334-4008 was used both for bandpass and flux calibration. For the execution performed in May, the same flux calibrator was used while J0522-3627 was used for bandpass calibration. For both execution blocks, J0348-2749 was used for phase calibration. 

The data have been reduced and calibrated using a standard pipeline running on Common Astronomy Software Application (\textsc{casa}) version 5.1.1-5 for the band 6 data taken in cycle 5, while a newer version (5.4.0-68) was used for the band 8 data. For the cycle 3 band 6 data, \textsc{casa} version 4.5.3 was used. Data from different execution blocks and the different cycles was combined before imaging. Imaging was done using the \textsc{casa} task \texttt{\small clean}. In the cleaning procedure, we first produced a dirty image from which the rms (root-mean-square) was measured. We verify that the measured rms in these images is consistent with the theoretically expected noise level. For the clean image, we set the cleaning threshold to 2 times the rms measured in the dirty image. 

Our imaging procedure is as follows: We create image cubes and continuum images using a natural weighting. For line images, we make cubes with several different spectral resolutions ranging from the native spectral resolution down to a resolution of \Edit{ $\sim 200 \ \mathrm{km\ s^{-1}}$ in both bands. Our synthesized beam in the band 6 continuum image is $0\farcs  87 \ \times \  0\farcs  67$, with a position angle of $\mathrm{P.A.}=-81^{\circ}$ . The corresponding values for the band 8 continuum image is $0\farcs  34 \ \times \  0\farcs  25$, $\mathrm{P.A.}=-88^{\circ}$. The synthesized beams in our [\ion{C}{ii}] and [\ion{O}{iii}] images are $0\farcs  89 \ \times \  0\farcs  69$, $\mathrm{P.A.}=-84^{\circ}$ and $0\farcs  36 \ \times \  0\farcs  26$, $\mathrm{P.A.}=89^{\circ}$, respectively. Given the apparent size of the source in the UV images (with an estimated diameter of $\sim 0\farcs3$) and the synthesized beam in band 8, we also create tapered images using several different tapering parameters up to $0\farcs 35$, which results in a beam major axis of $\approx 0\farcs  6$ in both line and continuum images, and an increase in the measured rms up to a factor $\sim 1.2$. We have also created [\ion{C}{ii}] image cubes with a tapering parameter of $0\farcs5$, since recent results indicate that the [\ion{C}{ii}]-emitting region may be 2 -- 3 times larger than the UV component \citep{carniani_kiloparsec-scale_2018, carniani_missing_2020}. This results in a beam major axis of $\approx 1\farcs1$, which should make sure we are covering over 3 times the UV size. This also leads to an increase in the noise level by a factor of $\sim 1.1$. We inspect the cubes and images taking into account the possible velocity offset between the FIR lines and $\mathrm{Ly\alpha}$, but none of the above strategies lead to significant detections of line nor continuum emission which could be associated with the target}. We have assessed the Gaussianity of the pixel noise distribution in all the images used to define upper limits on the non-detections in this paper. We adopted a 3$\sigma$ threshold as an upper limit for a non-detection in our analysis.


The flux calibration of ALMA data using \Edit{QSOs} (as is our case) relies on the frequent monitoring of a reference sample of QSOs in bands 3 and 7. This strategy provides an estimate of their intrinsic flux and spectral index and allows the extrapolation of their fluxes to the observed dates and frequencies. While the relative uncertainty on the flux calibration reported by the ALMA pipeline is generally below $\sim10\%$ for the data used here, the variability of the QSOs may introduce an additional uncertainty\footnote{ALMA Technical Handbook; \url{https://almascience.eso.org/documents-and-tools/cycle7/alma-technical-handbook}}. In order to get a handle on the flux uncertainty due to this variation, we extracted historical measurements of the flux calibrators from the ALMA source catalogue and used the maximum and minimum spectral index within a month around each observing date to estimate the uncertainty in the extrapolated flux density for our observing frequencies and dates. Given the monitoring intervals of our calibrators and the different criteria that can be established to estimate the spectral index (\Edit{e.g.} involved bands and contemporaneity of measurements) we consider this to be a conservative and robust estimate. For our band 8 observations, we find that the variability exhibited by J0522-3627 can lead to a difference in the derived flux density on the order of 30\%. Using a similar procedure for the band 6 observations, we find that the corresponding value is around 10\%.

\begin{table*}
\caption{Summary of ALMA observations. The table shows the date of observation, baseline lengths, number of antennae ($\mathrm{N_{ant}}$), central frequency of the SPW, integration time and median precipitable water vapor (PWV) for the band 6 and band 8 observations.}              
\label{table:ALMA_observations}      
\renewcommand{\arraystretch}{1.25}
\begin{center}                                  
\begin{tabular}{l l l l l l}          
\hline\hline                        
Date & Baseline lengths & $\mathrm{N_{ant}}$ & SPW central frequency & Integration time & Median PWV \\
YYYY-MM-DD & (m) & & (GHz) & (mm:ss) & (mm) \\
    \hline

    \multicolumn{6}{c}{Band 6}\\

    2018-04-19 & 15-500 & 42 & 219.35, 221.09, 233.98, 235.98 & 38:32 & 1.8 \\  
    2018-04-19 & 15-500 & 44 & 219.35, 221.09, 233.98, 235.98 & 38:30 & 1.8 \\
    2018-09-01 & 15-784 & 47 & 219.37, 221.11, 234.00, 236.00 & 38:33 & 1.3 \\
    
    \multicolumn{6}{c}{}\\
    
    2016-03-22\tablefootmark{a} & 15-460 & 40 & 219.00, 220.75 , 233.94, 235.82  & 20:47 & 2.8 \\  
    2016-05-15\tablefootmark{a} & 17-640 & 38 & 219.01, 220.77, 233.96, 235.83 & 20:47 & 1.6 \\
    
    \hline

    \multicolumn{6}{c}{Band 8}\\

    2018-08-25 & 15-500 & 43 & 391.64, 393.31, 403.71, 405.51 & 48:07 & 0.3 \\  
    2018-09-11 & 15-1200 & 48 & 391.63, 393.31, 403.70, 405.50 & 48:08 & 0.7 \\
    2018-09-11 & 15-1200 & 45 & 391.63, 393.31, 403.70, 405.50 & 48:07 & 0.7 \\
    2018-09-23 & 15-1400 & 45 & 391.63, 393.30, 403.70, 405.50 & 48:09 & 0.6 \\
    
\hline                                             
\vspace{8pt}
\end{tabular}
\end{center} 

\tablefoot{ \tablefoottext{a}{Cycle 3 observations; 2015.1.00821.S, PI: S. Finkelstein}}
\vspace{1pt}
\end{table*}

\section{Results}
\label{sec:results}
In the following sections, we present upper limits of the [\ion{O}{iii}] 88 $\mu \mathrm{m}$ and  [\ion{C}{ii}] 158 $\mu \mathrm{m}$ emission lines along with upper limits for the $87 \ \mu \mathrm{m}$ and $152 \ \mu \mathrm{m}$ dust continuum measurements, since no line or continuum emission is detected \Edit{(see Fig.~\ref{fig:ALMA_HST} and Fig.~\ref{fig:ALMA_spec})}. We also present results obtained by combining our ALMA upper limits and data in the rest-frame ultraviolet of z7\_GSD\_3811. \Edit{In the GOODS-S field, a small systematic offset between ALMA and HST images has been observed \citep[see e.g.][]{dunlop_deep_2017}. We have therefore used the coordinates from the Hubble Legacy Fields (HLF) GOODS-S catalog \citep{whitaker_hubble_2019}, which have been corrected using the Gaia DR2 catalog \citep{gaia_gaia_2016,gaia_gaia_2018}. We also use images from the HLF GOODS-S data release 2.0 (Illingworth et al. 2019, in prep). We compare the position of a foreground detected object (see Sect.~\ref{sec:dust_upperlimit}) and the HST image with the corrected coordinates and find that these are consistent within $\approx 0\farcs1$. }

\subsection{Upper limits for the [\ion{O}{iii}] and [\ion{C}{ii}] line fluxes}
\label{sec:Lines_upperlimit}

\begin{figure*}
    \centering
    \includegraphics[width=17cm]{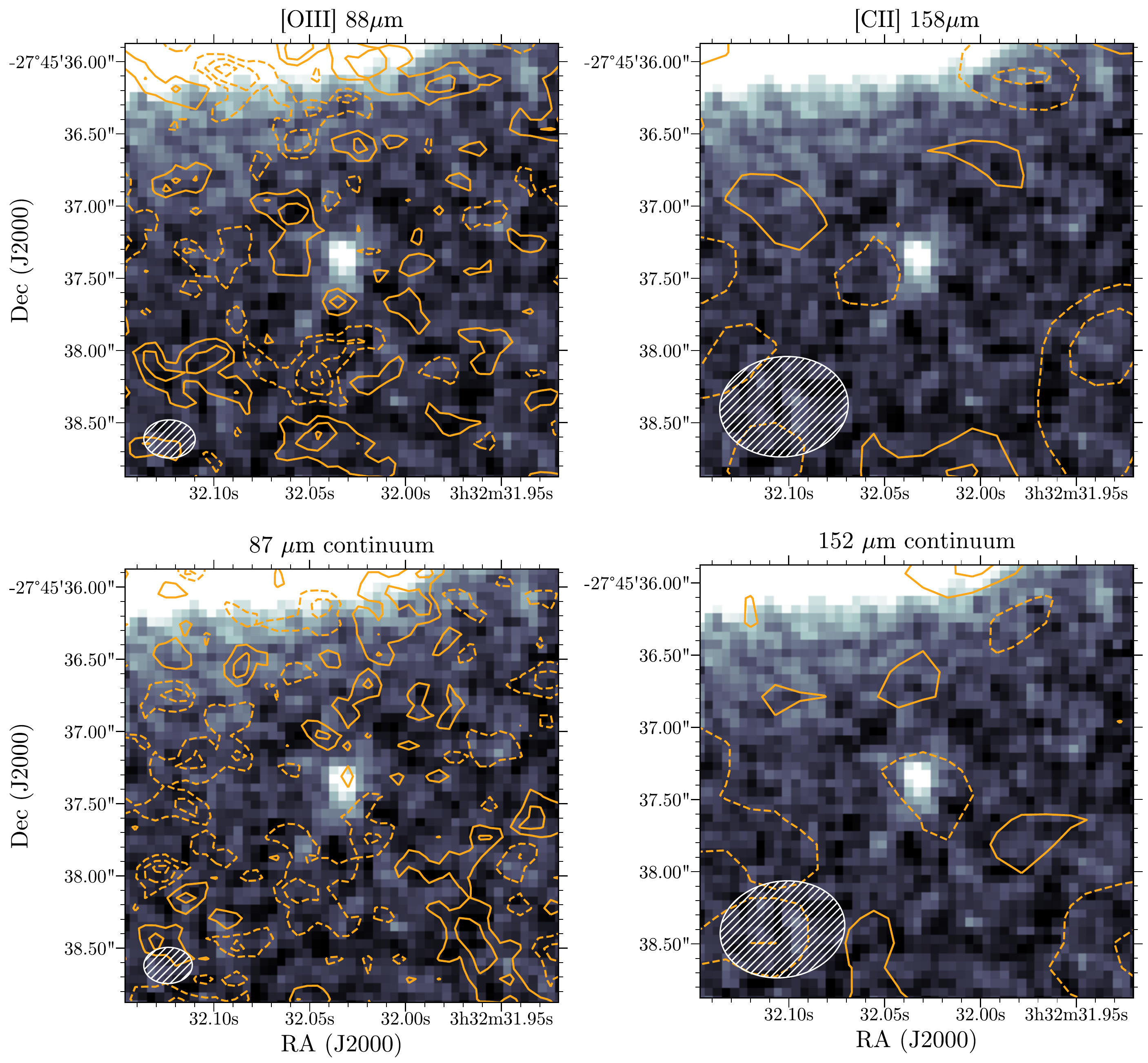}
\caption{ALMA observations drawn at $(-3,-2,-1,1,2,3)\ \times \sigma$ overlaid on the HST F160W image of z7\_GSD\_3811. The white hatched ellipse in the bottom of each frame shows the synthesized beam of ALMA, positive and negative contours are drawn with solid and dashed lines respectively. \textit{Top left:} [\ion{O}{iii}] line contours for an image collapsed over $ 400 \ \mathrm{km \ s^{-1}}$ around $391.63 \ \mathrm{GHz}$, where \Edit{$\sigma = 44 \ \mathrm{mJy\ km \ s^{-1} \ beam^{-1}}$.} \textit{Top right:} [\ion{C}{ii}] line contours for an image collapsed over $ 400 \ \mathrm{km \ s^{-1}}$ around $219.37 \ \mathrm{GHz}$, where $\sigma = 19 \ \mathrm{mJy\ km \ s^{-1} \ beam^{-1}}$. \textit{Bottom left:} Dust continuum contours at $87 \ \mu \mathrm{m}$, where $\sigma = 26 \ \mathrm{\mu Jy \ beam^{-1}}$. \textit{Bottom right:} Dust continuum contours at $152 \ \mu \mathrm{m}$, where $\sigma = 9.9 \ \mathrm{\mu Jy \ beam^{-1}}$.}
\label{fig:ALMA_HST}
\end{figure*}

\begin{figure}
    \centering
    \resizebox{\hsize}{!}{\includegraphics{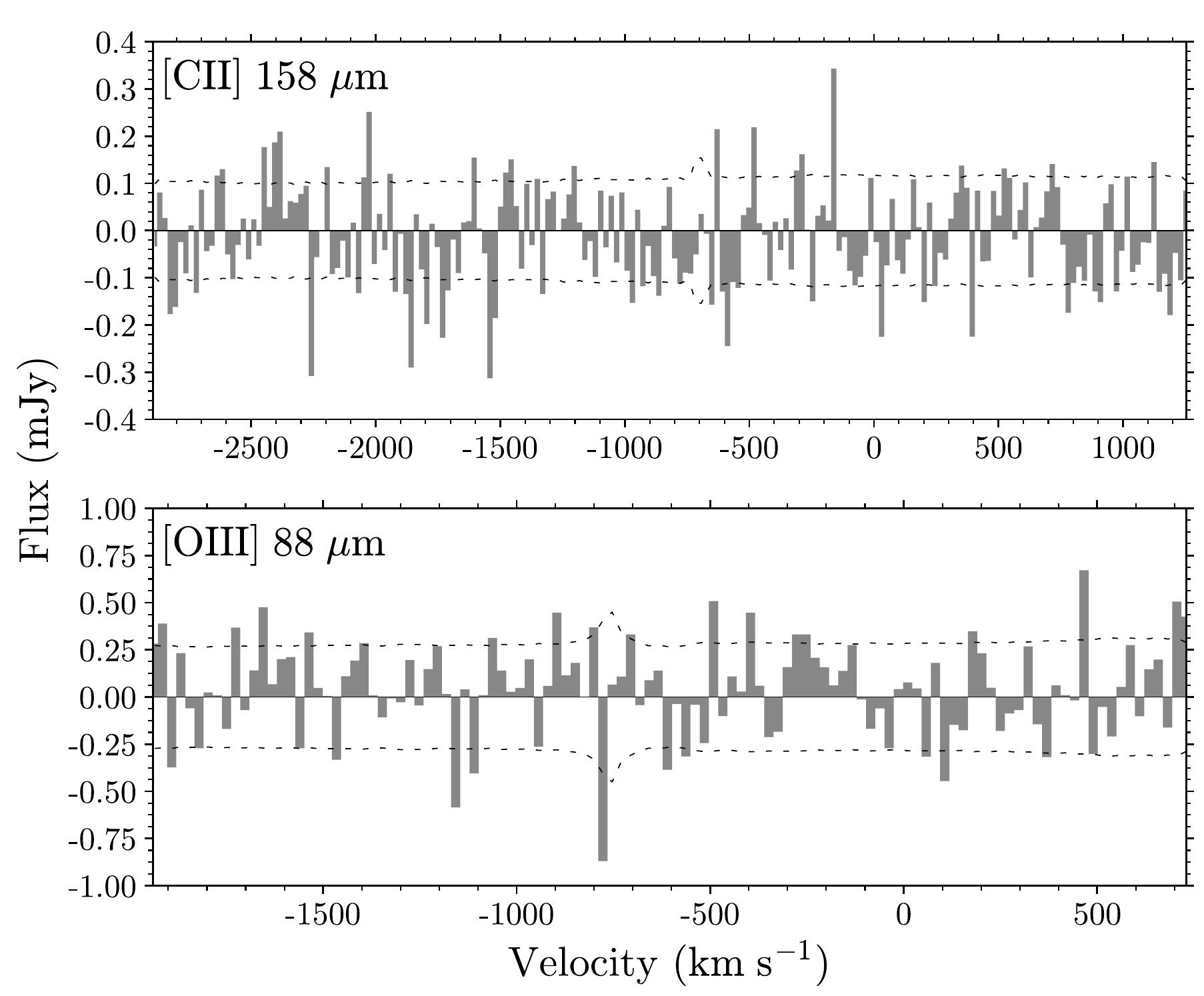}}
    \caption{\Edit{ [\ion{C}{ii}] (\emph{top}) and [\ion{O}{iii}] (\emph{bottom}) spectrum of z7\_GSD\_3811 shown at a resolution of $\approx 21 \ \mathrm{km \ s^{-1}}$ and $\approx 24 \ \mathrm{km \ s^{-1}}$, respectively. The spectra have been extracted in circular apertures with areas approximately equal to the beam areas. The horizontal axis shows the velocity relative to $\mathrm{Ly}\alpha$ and the dashed lines show the $1\sigma$ noise level obtained by randomly placing a large number of apertures over the image and adopting the variation in the flux as the noise.}}
    \label{fig:ALMA_spec}
\end{figure}

In order to search for [\ion{O}{iii}] and [\ion{C}{ii}] line-emission, we visually inspect the image cubes around the position of z7\_GSD\_3811 at different velocity slices relative to the expected position of the lines given the source redshift. For the [\ion{O}{iii}] line, we search for line emission between $\approx -2000 \text{ to } 700 \ \mathrm{km \ s^{-1}}$ relative to the expected position of the line ($391.63 \ \mathrm{GHz}$ at $z=7.6637$), while for [\ion{C}{ii}] the range that we use to search for line emission is $\approx -2900 \text{ to } 1300 \ \mathrm{km \ s^{-1}}$ relative to the expected position ($219.37 \ \mathrm{GHz}$ at $z=7.6637$). As mentioned in Sect.~\ref{sec:ALMA_b6_b8_obs}, the SPWs were arranged in this way in order to account for a possible velocity offset between the $\mathrm{Ly\alpha}$ redshift and the [\ion{O}{iii}] and [\ion{C}{ii}] lines. Note that since the cycle 3 and cycle 5 band 6 observations cover slightly different frequency ranges and have different spectral resolution, some edge channels are excluded in the imaging step. 

\Edit{ Given the lack of a clear detection from visual inspection, we also perform a systematic search for statistically significant features in our $\sim 50 \ (40), 100 \text{ and } 200\ \mathrm{km\ s^{-1}}$ cubes in band 8 (band 6), including tapered versions. Ideally, in order to claim a firm detection, we would want a signal at the $5\sigma$ level. Although, we also look for more marginal features that match the expected properties of the lines. In the search for potential signal, we select all pixels above 3 times the image rms within $1\arcsec$ of the target position. We analyse these further by visually inspecting spectra extracted at the locations and creating moment-0 maps around channels of high intensity. We also look at groups of pixels that are outside our $1\arcsec$ area, in cases where these give rise to any pixels above 3 times the image rms inside the area. In this procedure, we do not find any potential line emission signal with a peak signal-to-noise ratio (SNR) larger than 3 at the position of the UV component of z7\_GSD\_3811 (see also the [\ion{C}{ii}] and [\ion{O}{iii}] cube spectra extracted within one beam area at the position of the target in Fig.~\ref{fig:ALMA_spec}). In the [\ion{O}{iii}] cube, we are able to find features such that the collapsed moment maps contain some pixels that reach above 4 times the image rms within $\sim 1\arcsec$ of the position of z7\_GSD\_3811 (at distances of $\gtrsim 0\farcs5$). In addition to the spatial offset, these features are very narrow ($\mathrm{FWHM} \lesssim 40 \ \mathrm{km \ s^{-1}}$) compared to earlier [\ion{O}{iii}] detections and exhibit large velocity offsets relative to $\mathrm{Ly\alpha}$ ($\gtrsim 600 \ \mathrm{km \ s^{-1}}$), and thus are unlikely to belong to the [\ion{O}{iii}] line. Similarly, we find two marginal features at a $\sim 3.6-3.9\sigma$ level around the target position in the [\ion{C}{ii}] cube that also exhibit significant spatial ($\gtrsim0\farcs8$) and spectral ($\gtrsim 700 \ \mathrm{km \ s^{-1}}$) offsets. In most of our collapsed maps, we are also able to find collections of negative pixels at the same significance as the positive ones when inspecting a larger region of the image, and therefore cannot reliably distinguish these features from noise. We also inspect the pixel noise distributions in the collapsed images to see if there are any clear deviations from Gaussian noise. For the brightest features, we calculate an integrated SNR by fitting a 2D-Gaussian using the \textsc{casa} task \texttt{\small imfit} and separately image the data taken at different observation times in order to see if the signal is consistent within these.} 

\Edit{We find a very narrow $\mathrm{FWHM}\approx 30 \ \mathrm{km \ s^{-1}}$, at $\approx -1250\ \mathrm{km \ s^{-1}}$ relative to $\mathrm{Ly\alpha}$ ) potential signal at $\approx 1\farcs 2$ to the south-west of the target in band 8. When imaged with a beam of $0\farcs  47 \ \times \  0\farcs  38$ (corresponding to a $0\farcs 2$ taper), and collapsed over $\approx 70\ \mathrm{km \ s^{-1}}$, the peak pixel reaches 5 times the image rms. However, the integrated SNR given by \texttt{\small imfit} is only $\approx 3.1$. Given the large spectral and spatial offset relative to the UV component of z7\_GSD\_3811 and the narrow line-width, this possible signal is unlikely to belong to our target. We also note that there are some positive channels around $-240\ \mathrm{km \ s^{-1}}$ in the [\ion{O}{iii}] spectrum shown in Fig.~\ref{fig:ALMA_spec}. When fitted with a gaussian, we find a line-width around $ 100 \ \mathrm{km \ s^{-1}}$. By optimally selecting channels to collapse in order to maximize the SNR, we find that one pixel reaches a SNR of 3 at the eastern edge of the UV component, $\approx 0.1\arcsec$ from the center. This feature has a line-width and velocity offset which is consistent with expected values for [\ion{O}{iii}], however, given its very marginal significance, we are unable to identify this as the line. After all these checks, we are unable to find any potential [\ion{C}{ii}] and [\ion{O}{iii}] emission that can be reliably distinguished from noise and which can be associated with our target. Note also that none of the potential emission features found within $1\arcsec$ exhibit integrated SNRs above 3. Since we are unable to detect any of the targeted lines, we proceed to derive upper limits.}

\Edit{As mentioned in Sect.\ref{sec:target}, z7\_GSD\_3811 has a detected emission line which is consistent with $\mathrm{Ly\alpha}$ at $z=7.6637$. There is, however, a possibility that this line belongs to the [\ion{O}{ii}] $\lambda 3727$ doublet at $z\approx1.8$. This scenario is discussed by \citet{song_keck_2016}, who, while unable to rule out a low-redshift interpretation, argue that the asymmetric line-profile, the photometric redshift determination and non-detection in optical (and stacked optical) bands in combination with their SED-fitting suggest that the detected line is $\mathrm{Ly\alpha}$ at a redshift of $z\approx 7.7$.} 

In order to make a reasonable assumption on the line-width expected for the [\ion{C}{ii}] and [\ion{O}{iii}] lines, we look at other high-redshift objects. In the case of [\ion{C}{ii}], there are several successful detections in the high-redshift Universe \citep[see e.g.][for a summary]{matthee_resolved_2019,harikane_large_2019}. Looking at objects with similar UV luminosities as z7\_GSD\_3811, typical line-widths are ${v_{\mathrm{FWHM,[\ion{C}{ii}]}} \approx 50 \text{ -- } 150 \ \mathrm{km \ s^{-1}}}$ \citep{pentericci_tracing_2016, carniani_kiloparsec-scale_2018}. In the case of [\ion{O}{iii}]-detected objects with similar UV luminosities, these exhibit line-widths of ${v_{\mathrm{FWHM,[\ion{O}{iii}]}} = 80 \text{ -- } 150 \ \mathrm{km \ s^{-1}}}$ \citep{inoue_detection_2016, tamura_detection_2019}. Thus, as our fiducial line width for both the [\ion{C}{ii}] and [\ion{O}{iii}] line width, we use $100 \ \mathrm{km \ s^{-1}}$. \Edit{Considering the luminosities that we derive from our upper limits, this [\ion{C}{ii}] line-width is also in good agreement with values presented in \citet{kohandel_kinematics_2019} and \citet{schaerer_ALPINE_2020}}. However, throughout the paper, we also show the results that are obtained if one assumes a 4 times larger line-width, which is similar to the $\mathrm{Ly\alpha}$ line-width \citep{song_keck_2016}. In order to estimate an upper limit to the line fluxes, we thus calculate the integrated rms in a collapse of $\Delta v = 100$ and $ 400 \ \mathrm{km \ s^{-1}}$ around the expected positions of the lines given the source redshift ($391.63$ and $219.37 \ \mathrm{GHz}$ for the [\ion{O}{iii}] and [\ion{C}{ii}] line, respectively). Note that it is likely that the $\mathrm{Ly\alpha}$ line exhibits a velocity offset relative to the actual systemic redshift \citep[see e.g.][]{hashimoto_big_2019}. However, with the exception of an atmospheric line around $-800 \ \mathrm{km \ s^{-1}}$ in band 8 and an increased rms around $-700 \ \mathrm{km \ s^{-1}}$ in band 6 due to some completely flagged channels in the cycle 3 data, our rms is relatively invariant in the observed range. Between $v = -500 \ \mathrm{km \ s^{-1}}$ and $v = 250 \ \mathrm{km \ s^{-1}}$, the variation in the integrated rms for a $100 \ \mathrm{km \ s^{-1}}$ wide line can be up to approximately $\pm 2\%$ in band 8 and $+2\%, -4\%$ in band 6. Thus, different assumptions on the line centers should not alter our results significantly.

The measured rms in our collapsed [\ion{O}{iii}] images is \Edit{$ 23 $ and $44 \ \mathrm{mJy\ km \ s^{-1} \ beam^{-1}}$} for $\Delta v = 100$ and $ 400 \ \mathrm{km \ s^{-1}}$, respectively. For the corresponding [\ion{C}{ii}] images, we measure an rms of $ 9.8 $ and $19 \ \mathrm{mJy\ km \ s^{-1} \ beam^{-1}}$, respectively. Continuum-subtracting the spectrum would lead to an increase in the rms in the \Edit{channels excluded from the fitting} \citep[e.g.][]{sault_analysis_1994}, thus depending on the assumed line width. Since our spectrum is compatible with zero, we proceed without performing continuum subtraction. \Edit{We show collapsed line images and spectra in Fig.~\ref{fig:ALMA_HST} and Fig.\ref{fig:ALMA_spec}, respectively}.

Assuming that the target is unresolved with the native beam sizes, these fluxes correspond to $3\sigma$ upper limits on the line luminosities of \Edit{$L_{\mathrm{[\ion{O}{iii}]}} < 1.6 \ \times \ 10^{8} \  \mathrm{L_{\odot}}$, $L_{\mathrm{[\ion{O}{iii}]}} < 3.2 \ \times \ 10^{8} \  \mathrm{L_{\odot}}$} for the [\ion{O}{iii}] line assuming a line-width of $\Delta v = 100$ and $ 400 \ \mathrm{km \ s^{-1}}$, respectively. The corresponding values for [\ion{C}{ii}] are $L_{\mathrm{[\ion{C}{ii}]}} < 4.0 \ \times \ 10^{7} \ \mathrm{L_{\odot}}$, $L_{\mathrm{[\ion{C}{ii}]}} < 7.5 \ \times \ 10^{7} \ \mathrm{L_{\odot}}$. \Edit{Note that since the rms scales as the square-root of the line-width, the $3\sigma$ upper-limits obtained for $ 400 \ \mathrm{km \ s^{-1}}$ are comparable to the value one would obtain for a $6\sigma$ upper limit assuming a $100 \ \mathrm{km \ s^{-1}}$ line}.

\subsection{Upper limits for the $87 \ \mu \mathrm{m}$ and $152 \ \mu \mathrm{m}$ dust continuum}
\label{sec:dust_upperlimit}
In order to create dust continuum images, we combine all four SPWs in each band and create collapsed images around $87 \ \mu \mathrm{m}$ and $152 \ \mu \mathrm{m}$ for the band 8 and band 6 data respectively. Since no significant line emission is detected (see Sect.~\ref{sec:Lines_upperlimit}), we use the full frequency ranges in each band. While no FIR continuum is detected at the location of z7\_GSD\_3811, a foreground continuum source is detected in both the band 6 and band 8 dust continuum images\footnote{The position of the foreground object is: R.A.$=$~03:32:32.91~Dec.~$=$~-27:45:41.01 (J2000), outside the zoomed-in images shown in Fig.~\ref{fig:ALMA_HST}. \Edit{The position of this source overlaps with an object in the HST F160W image.}}. Before measuring the rms in the continuum images, we mask out the detected foreground object. Our measured rms in the $87 \ \mu \mathrm{m}$ image is $26 \ \mathrm{\mu Jy \ beam^{-1}}$. In the $152 \ \mu \mathrm{m}$ image, we measure an rms of $9.9 \ \mathrm{\mu Jy \ beam^{-1}}$. The continuum images are shown in Fig.~\ref{fig:ALMA_HST}. Note that masking out channels corresponding to $ 400 \ \mathrm{km \ s^{-1}}$ around the expected positions of the line centers from the continuum images leads to an insignificant increase in the measured rms ($\leq 3\%$).  

We estimate the total infrared (IR) dust luminosity ($L_{IR}$) by integrating a modified black-body \citep{DeBreuck_emission_2003} curve over 8 -- 1000 $\mu \mathrm{m}$ while assuming an emissivity index of $\beta_d=1.5$ and a dust temperature of $T_d = 45 \ \mathrm{K}$. These values are largely consistent with assumptions made in other studies \citep[e.g.][]{ouchi_intensely_2013, Schaerer_constraints_2015, inoue_detection_2016, hashimoto_onset_2018, matthee_resolved_2019} and to that observed in \citet{knudsen_merger_2017}. Observations of local galaxies suggest that these have slightly lower dust temperatures \citep[$\approx 20 \text{ -- } 40 \ \mathrm{K}$; ][]{remy-ruyer_revealing_2013}. On the other hand, recent studies suggest that the dust temperatures in high-redshift galaxies may be significantly higher \citep{bakx_ALMA_2020}. In order to get a handle on the variation in $L_{IR}$ for different assumptions regarding the dust temperature we also estimate $L_{IR}$ assuming dust temperatures of $35 \text{ and } 55 \ \mathrm{K}$. In this procedure, we correct for dust heating and the background effect of the CMB following \citet{Ota_ALMA_2014}. Note that the correction factor for CMB effects at $87 \ \mu \mathrm{m}$ when assuming $T_d \ = \ 45 \ \mathrm{K}$ is basically unity, while a value of $\approx 0.88$ is found for the $152 \ \mu \mathrm{m}$ dust continuum. The obtained $3\sigma$ upper limits to the total IR dust luminosities around $87 \ \mu \mathrm{m}$ and $152 \ \mu \mathrm{m}$ are $L_{\mathrm{IR,87\mathrm{\mu m},45K}} \ <9.3 \ \times  \ 10^{10} \ \mathrm{L_{\odot}}$ and $L_{\mathrm{IR,152\mathrm{\mu m},45K}} <9.1 \ \times  \ 10^{10} \ \mathrm{L_{\odot}}$, respectively. We will use these values in the forthcoming analysis (see table~\ref{table:Results}). For comparison, if we instead assume a dust temperature of 35 K or 55 K, we find: $L_{\mathrm{IR,87\mathrm{\mu m},35K}} \ <6.7 \ \times  \ 10^{10} \ \mathrm{L_{\odot}}$, $L_{\mathrm{IR,152\mathrm{\mu m},35K}} <4.9 \ \times  \ 10^{10} \ \mathrm{L_{\odot}}$, $L_{\mathrm{IR,87\mathrm{\mu m},55K}} \ <1.4 \ \times  \ 10^{11} \ \mathrm{L_{\odot}}$ and $L_{\mathrm{IR,152\mathrm{\mu m},55K}} <1.7 \ \times  \ 10^{11} \ \mathrm{L_{\odot}}$ respectively.

\begin{table}
\caption{Summary of properties of z7 GSD 3811 observed with/ derived from HST and ground-based imaging, ground-based spectroscopy and ALMA observations.}
\label{table:Results}      

\renewcommand{\arraystretch}{1.25}

\centering                                      
\begin{tabular}{l l}          
\hline\hline  
Property & \\

\hline      

R.A. (J2000)\tablefootmark{a} & 3:32:32.03 \\
Dec. (J2000)\tablefootmark{a} & -27.45:37.1 \\
$z_{\mathrm{Ly\alpha}}$ \tablefootmark{b} & $7.6637 \pm 0.0011$ \\
$M_{\mathrm{UV}}$ \tablefootmark{b} & $-21.22_{-0.10}^{+0.06}$ \\ 
 & \\
$\mathrm{SFR_{UV}}  \quad (\mathrm{M_{\odot}} \ \mathrm{yr^{-1}}) $ \tablefootmark{c} & $12_{-1}^{+1}$ \\ 
 & \\
$S_{\nu,87\mu m} \quad (\mathrm{\mu Jy \ beam^{-1}})$ & $< 26 \ (1\sigma)$ \\
$S_{\mathrm{\nu,[\ion{O}{iii}]}} \Delta v \quad (\mathrm{mJy\ km \ s^{-1} \ beam^{-1}})$ \tablefootmark{d} & \Edit{$< 23 \ (1\sigma)$ }\\
$S_{\nu,152\mu m} \quad (\mathrm{\mu Jy \ beam^{-1}})$ & $< 9.9 \ (1\sigma)$ \\
$S_{\mathrm{\nu,[\ion{C}{ii}]}} \Delta v  \quad (\mathrm{mJy\ km \ s^{-1} \ beam^{-1}})$ \tablefootmark{d} & $< 9.8 \ (1\sigma)$ \\
 & \\
$L_{\mathrm{IR,87\mathrm{\mu m},45K}} \quad ( 10^{10} \ \mathrm{L_{\odot}})$ & $<9.3 \ (3\sigma)$ \\
$L_{\mathrm{IR,152\mathrm{\mu m},45K}} \quad ( 10^{10} \ \mathrm{L_{\odot}})$ & $<9.1 \ (3\sigma)$ \\

\Edit{$\Delta L_{\mathrm{IR,87\mathrm{\mu m},+10K}} \quad \mathrm{(dex)}$\tablefootmark{e}}  & \Edit{$0.18$} \\
\Edit{$\Delta L_{\mathrm{IR,152\mathrm{\mu m},+10K}} \quad \mathrm{(dex)}$\tablefootmark{e}}  & \Edit{$0.27$} \\
 & \\

$L_{\mathrm{[\ion{O}{iii}]}} \quad ( 10^{8} \ \mathrm{L_{\odot}})$ \tablefootmark{d} & \Edit{$<1.6 \ (3\sigma)$ }\\
$L_{\mathrm{[\ion{C}{ii}]}} \quad ( 10^{7} \ \mathrm{L_{\odot}})$ \tablefootmark{d} & $<4.0 \ (3\sigma)$ \\
 & \\
$\mathrm{SFR_{IR,45K}} \quad (\mathrm{M_{\odot}} \ \mathrm{yr^{-1}})$ \tablefootmark{c} & $<14 \ (3\sigma)$ \\ 
 & \\
\hline

\end{tabular}

\vspace{20pt}

\tablefoot{
\tablefootmark{a}{\Edit{Coordinates from \citet{song_keck_2016}, not corrected for offset between Gaia and HST.}}
\tablefoottext{b}{Values from \citet{song_keck_2016}}
\tablefoottext{c}{Value obtained using the conversion in \citet{kennicutt_star_2012} which uses a \citet{kroupa_variation_2001} initial mass function.}
\tablefoottext{d}{Values obtained by collapsing a spectral cube over $\Delta v = 100 \ \mathrm{km \ s^{-1}}$.} \tablefoottext{e}{\Edit{Change in  $L_{\mathrm{IR}}$ in the case of a $10 \mathrm{K}$ higher dust temperature.}}
}

\end{table}

\subsection{The [\ion{C}{ii}]-SFR relation}
\label{sec:CII_SFR}

As mentioned in Sect.~\ref{sec:introduction}, several studies have found that [\ion{C}{ii}] emission in high-redshift objects often is weak compared to local galaxies with similar SFRs. Using the UV magnitude and the relationship between UV luminosity and SFR by \citet{kennicutt_star_2012}, we find a UV SFR for z7\_GSD\_3811 of $\mathrm{SFR_{UV}} \ = \ 12_{-1}^{+1} \ \mathrm{M_{\odot}} \ \mathrm{yr^{-1}}$. Here, we use the UV magnitude from \citet{song_keck_2016} presented in table~\ref{table:Results}, and assume that the error bars are symmetric and equal to the larger error presented there. Whereas the UV luminosity gives us an estimate of the non-obscured SFR, the IR luminosity can be converted to an estimate of the dust-obscured star formation. Using the corresponding relationship for the IR luminosity \citep{kennicutt_star_2012} and the upper limit on the total IR luminosity obtained via the band 6 ALMA data, we derive an IR SFR of $\mathrm{SFR_{IR}} \ < 4.5 \ \mathrm{M_{\odot}} \ \mathrm{yr^{-1}}$ ($1\sigma$) for a dust temperature of 45 K.

\begin{figure}
    \centering
    \resizebox{\hsize}{!}{\includegraphics{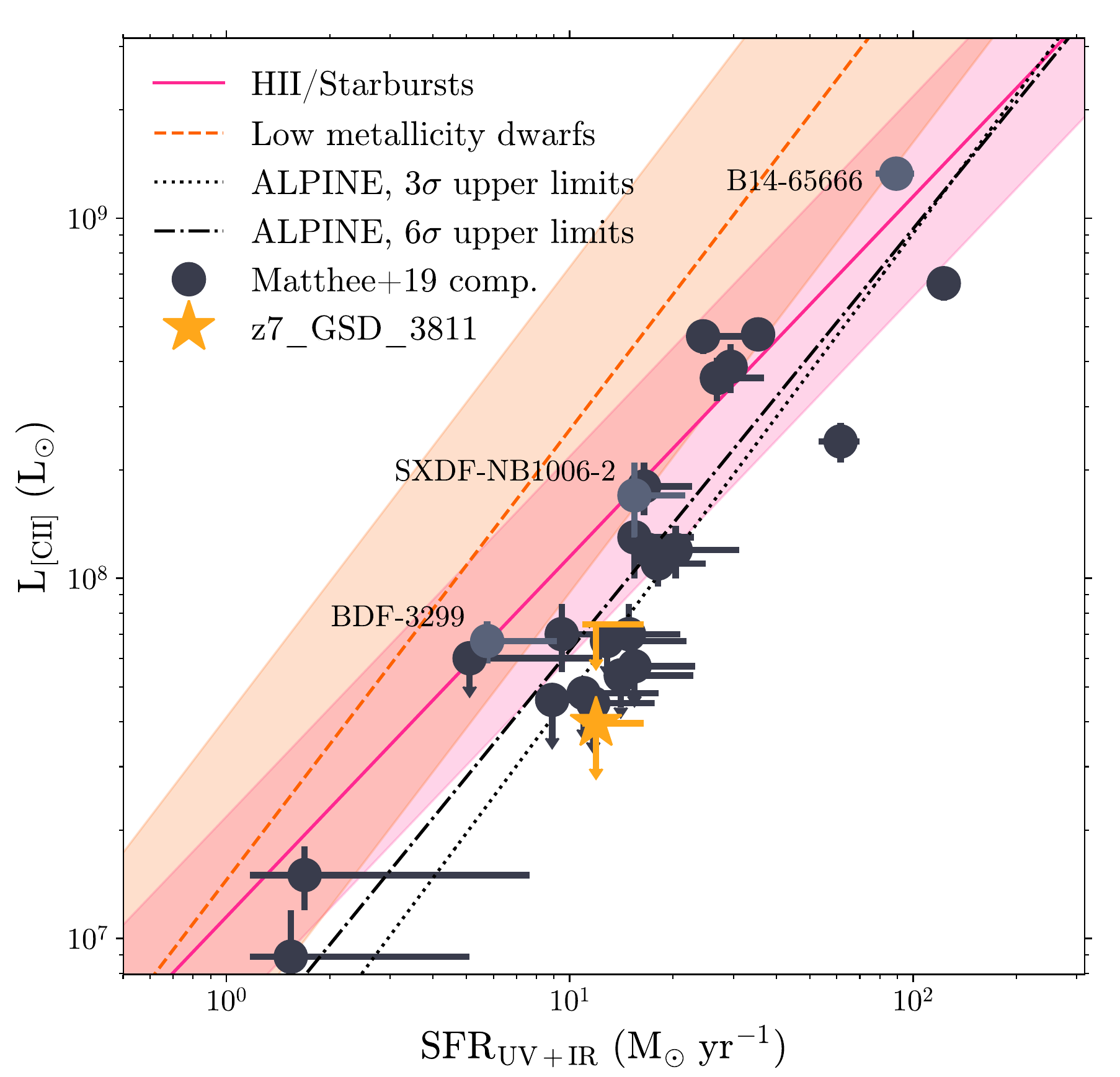}}
    \caption{[\ion{C}{ii}] $158 \ \mu \mathrm{m}$ luminosities versus the UV and IR SFR. The yellow star shows the upper limit of z7\_GSD\_3811 assuming a line-width of $100 \ \mathrm{km \ s^{-1}}$, while the dash and arrow show the upper limit obtained if we assume a line-width of $400 \ \mathrm{km \ s^{-1}}$. The pink line and orange dashed line show the relation for local HII/starburst galaxies and metal-poor dwarf galaxies by \citet{delooze_applicability_2014}, where the shaded regions correspond to the $1\sigma$ dispersion in the relations. \Edit{The dotted and dash-dotted lines show the relations from the ALPINE-ALMA survey, derived by \citet[][see text]{schaerer_ALPINE_2020}. The dark-gray circles show objects from the compilation by \citet{matthee_resolved_2019} with $z>6$. For BDF-3299 and SXDF-NB1006-2, we show updated [\ion{C}{ii}] measurement from \citep{carniani_missing_2020}}.  For this sample, we have re-scaled the upper limits to line-widths of $100 \ \mathrm{km \ s^{-1}}$. A few galaxies from the compilation that are also detected in [\ion{O}{iii}] are indicated by their names and a lighter gray color. All upper limits are at the $3\sigma$ level.}
    \label{fig:SFR_CII}
\end{figure}

Following \citet{matthee_resolved_2019}, we use this $1\sigma$ upper-limit to the IR SFR as our upper error bar to the UV+IR SFR ($\mathrm{SFR_{UV+IR}} \ = \ 12_{-1}^{+5} \ \mathrm{M_{\odot}} \ \mathrm{yr^{-1}}$). In order to see how our values compare to other high-redshift detections, we compare our derived UV+IR SFR and upper limit to the [\ion{C}{ii}] luminosity to galaxies at \Edit{$z>6$} from the recently published compilation by \citet{matthee_resolved_2019}. We derive IR and UV SFRs for the galaxies included in the compilation in the same way as for z7\_GSD\_3811 using the UV magnitudes and IR luminosities presented in \citet{matthee_resolved_2019}. While this compilation does have a UV magnitude for BDF-3299, we derive a value using the $\mathrm{Ly\alpha}$ and intergalactic medium (IGM) -corrected Y band observation and inferred UV-slope from \citet{vanzella_spectroscopic_2011}. We find $M_{UV}\approx -20.4$ for this object, which is $0.5$ magnitudes fainter than the original value in the compilation\footnote{\label{MUV_note} $M_{UV}\approx -20.4$ leads to a UV SFR that is about a factor of 0.65 lower than what we would obtain using $M_{UV}\approx -20.9$ assuming the \citet{kennicutt_star_2012} SFR relation. The value we derive is consistent with the UV SFR presented in \citet{maiolino_assembly_2015}. While there are some spatial offsets between different emission components in BDF-3299, we assume that the UV, [\ion{O}{iii}] and [\ion{C}{ii}] emission are associated with the same object.}. In the same way as for z7\_GSD\_3811, we use upper limits on the IR SFR as upper limits to the UV+IR SFR. While we derive UV and IR SFRs for all objects in a consistent manner, we also note that several objects in the compilation have published SFRs from SED fitting. These can in certain cases be significantly larger (or smaller) than the values we obtain using the method described here. For non-detections in [\ion{C}{ii}], we have gone back to the individual publications and re-scaled all the upper limits to the same line-width as z7\_GSD\_3811 (i.e. $100 \  \mathrm{km \ s^{-1}}$) by assuming that the velocity-integrated rms scales as the square root of the line width. \Edit{For BDF-3299 and SXDF-NB1006-2, we have used the updated measurements from \citet{carniani_missing_2020}.} In Fig.~\ref{fig:SFR_CII}, we show the position of z7\_GSD\_3811 and the compilation objects in the [\ion{C}{ii}]-SFR plot. We also show the local relation for HII/Starburst galaxies and metal-poor dwarf galaxies from \citet{delooze_applicability_2014}. The shaded region in the figure shows the $1\sigma$ dispersion in the relations. \Edit{Note that the relation for metal-poor dwarf galaxies is extrapolated to higher SFR}.

\Edit{Our upper limit on the [\ion{C}{ii}] line places z7\_GSD\_3811 below the local relations by \citet{delooze_applicability_2014}. As can be seen in Fig.~\ref{fig:SFR_CII}, several other objects with similar SFR also fall below the local relations. As is the case for z7\_GSD\_3811, many of these are undetected in [\ion{C}{ii}]. Even if we assume a [\ion{C}{ii}] line which is 4 times wider, our upper limit still places z7\_GSD\_3811 below the local relations, although with a smaller offset. The dotted line in Fig.~\ref{fig:SFR_CII} shows the relation by \citet{schaerer_ALPINE_2020}, where they have used detections and $3\sigma$ upper limits from the ALPINE-ALMA survey in combination with earlier $z>6$ measurements. \citet{schaerer_ALPINE_2020} correct for hidden star formation for those objects that are undetected in the IR by using the average IRX-$\beta$ relationship by \citet{fudamoto_ALPINE_2020}. Our upper limit on the [\ion{C}{ii}] luminosity places z7\_GSD\_3811 at a position which is below this relation in case we assume a $100 \ \mathrm{km \ s^{-1}}$, although, only marginally so when we consider uncertainties related to the fits of \citet{schaerer_ALPINE_2020}. For the assumption of a wider line, the upper limit is placed just above the relation. \citet{schaerer_ALPINE_2020} find that using more conservative upper limits ($6\sigma$) for the non-detections in their sample still leads to a relation which is somewhat steeper than the local one (the dash-dotted line in Fig~\ref{fig:SFR_CII}). Our upper limit on the [\ion{C}{ii}] luminosity could in principle make z7\_GSD\_3811 consistent with this relation under the assumption of a $400 \ \mathrm{km \ s^{-1}}$ line-width. This would also be the case if we used a $6\sigma$ upper limit for our $100 \ \mathrm{km \ s^{-1}}$ line.}

\subsection{IRX-$\beta$}
\label{sec:IRX-beta}

Dust-correcting rest-frame UV observations of galaxies requires making assumptions regarding the dust attenuation curve, or in other words, how a given amount of dust affects the shape of the spectrum. Since the heated dust emits at IR wavelengths, one way to try to constrain the dust-attenuation curve in galaxies is to look at the relation between the infrared excess (IRX; the IR-to-UV luminosity ratio) and the UV continuum slope ($\beta$). While no clear consensus has been reached regarding the dust attenuation curve at higher redshifts, there are studies that seem to favor a steep attenuation law similar to that of the Small Magellanic cloud (SMC) over a flatter \citet{calzetti_dust_2000} law \citep[e.g.][]{reddy_star_2006, capak_galaxies_2015, bouwens_ALMA_2016}.

\begin{figure}
    \centering
    \resizebox{\hsize}{!}{\includegraphics{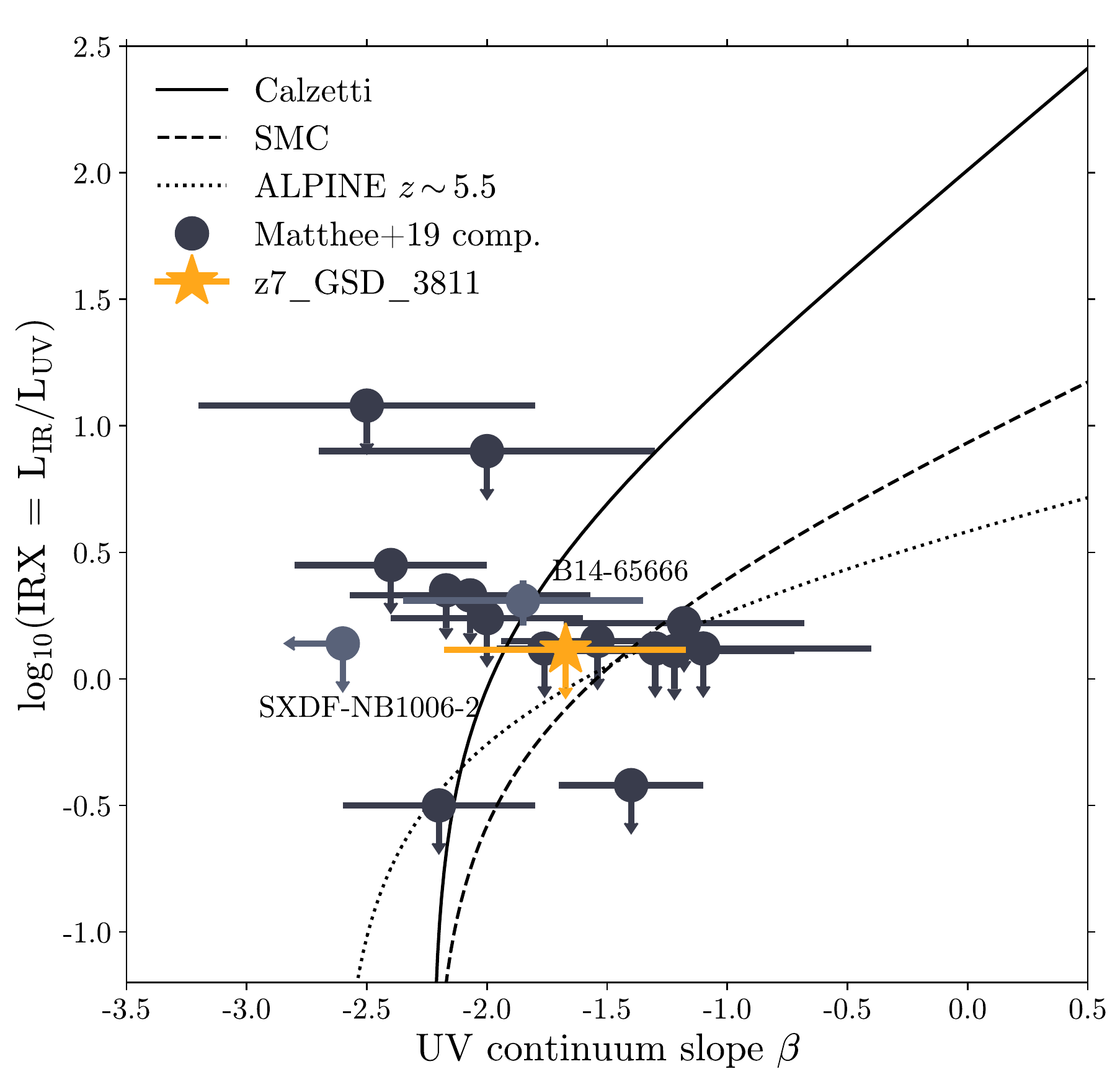}}
    \caption{The relation between the IR-to-UV luminosity ratio (IRX) and the UV slope $\beta$. The yellow star shows the upper limit on the IRX calculated from the band 6 observations z7\_GSD\_3811. We also show the IRX and UV slopes of galaxies from the compilation of \citet[][dark-gray circles]{matthee_resolved_2019}. IRX values are calculated under the assumption of a dust temperature of 45K and a dust emissivity index of $\beta_{\mathrm{d}}=1.5$. The solid and dashed line represent the Calzetti and SMC dust laws calculated following \citet{bouwens_ALMA_2016}. \Edit{The dotted line shows the IRX-$\beta$ relation from the ALPINE-ALMA survey data, derived by \citet[][see text]{fudamoto_ALPINE_2020}}. Galaxies from the \citet{matthee_resolved_2019} compilation that are also detected in [\ion{O}{iii}] are indicated by their names and a lighter gray color. Upper limits on the IRX are at the $3\sigma$ level.}
   \label{fig:IRX_Beta}
\end{figure}

We calculate the position of z7\_GSD\_3811 in the IRX-$\beta$ plane and compare it to the positions of other high-redshift objects (\Edit{$z>6$}) from the compilation by \citet{matthee_resolved_2019} in Fig.~\ref{fig:IRX_Beta}. Following \citet{ono_stellar_2010}, we find a UV slope for z7\_GSD\_3811 using the F125W and F160W fluxes available in the official CANDELS catalog \citep{guo_candels_2013}. We obtain F125W$=25.89_{-0.07}^{+0.08}$, F160W$=25.81_{-0.08}^{+0.09}$ and $\beta \approx -1.7\pm 0.5$. Note that the magnitudes derived from the CANDELS catalog differ slightly from the ones presented in \citet{song_keck_2016}. If we would use the values presented there ($25.8$, $25.9$, for the two filters, respectively) we would obtain a significantly bluer UV slope $\beta \approx -2.4$. Indeed, \citet{song_keck_2016} find that their best fit SED also exhibits a significantly bluer UV slope $\beta = -2.2^{+0.3}_{-0.2}$. However, considering the large uncertainties related with these UV slopes, these values are largely consistent with each other. In Fig.~\ref{fig:IRX_Beta}, we use $\beta \approx -1.7\pm 0.5$. For the calculation of the IRX, we use the $3\sigma$ upper limit on the IR dust luminosity from the band 6 observations, a dust temperature of 45 K and a dust emissivity index of $\beta_{\mathrm{d}}=1.5$ in order to be consistent with the IR luminosities presented in \citet{matthee_resolved_2019}. For both the literature sample and z7\_GSD\_3811, UV luminosities are calculated as $L_{\mathrm{UV}}=\nu L_{\nu}$ at $1500 \ \mathrm{\AA}$ rest-frame, where $L_{\nu}$ is the rest-frame luminosity in units of $\mathrm{erg\ Hz^{-1} \ s^{-1}}$ and $\nu$ is the frequency corresponding to $1500 \ \mathrm{\AA}$. For z7\_GSD\_3811, we use the UV magnitude from \citet{song_keck_2016} presented in table~\ref{table:Results}. We obtain $\mathrm{log_{10}(IRX)} \lesssim 0.11$ (see Fig.~\ref{fig:IRX_Beta}). 

\Edit{In Fig.~\ref{fig:IRX_Beta}, we show the IRX-$\beta$ relations of the Calzetti and SMC dust laws calculated following \citet{bouwens_ALMA_2016} assuming an intrinsic UV slope $\beta_0=-2.23$. We also show the recently derived relation for $z\sim 5.5$ galaxies from the ALPINE-ALMA survey \citep{fudamoto_ALPINE_2020},  which is even steeper than the SMC relation. The ALPINE-ALMA relation uses an intrinsic UV slope $\beta_0=-2.62$ derived by \citet{reddy_HDUV_2018} based on the binary population and spectral synthesis models \citep[BPASS;][]{eldridge_effect_2012,stanway_stellar_2016}. Our upper limit on the IRX in z7\_GSD\_3811 is largely consistent with either of the three relations. While the related uncertainties are large, the bluer UV slope obtained by \citet{song_keck_2016} would place z7\_GSD\_3811 above all three relations. Deeper dust continuum observations and future spectroscopic measurements of the UV slope in z7\_GSD\_3811 will likely give more insight into which of these relations best describes the object.}

\subsection{[\ion{O}{iii}] and UV luminosity}
\begin{figure}
    \centering
    \resizebox{\hsize}{!}{\includegraphics{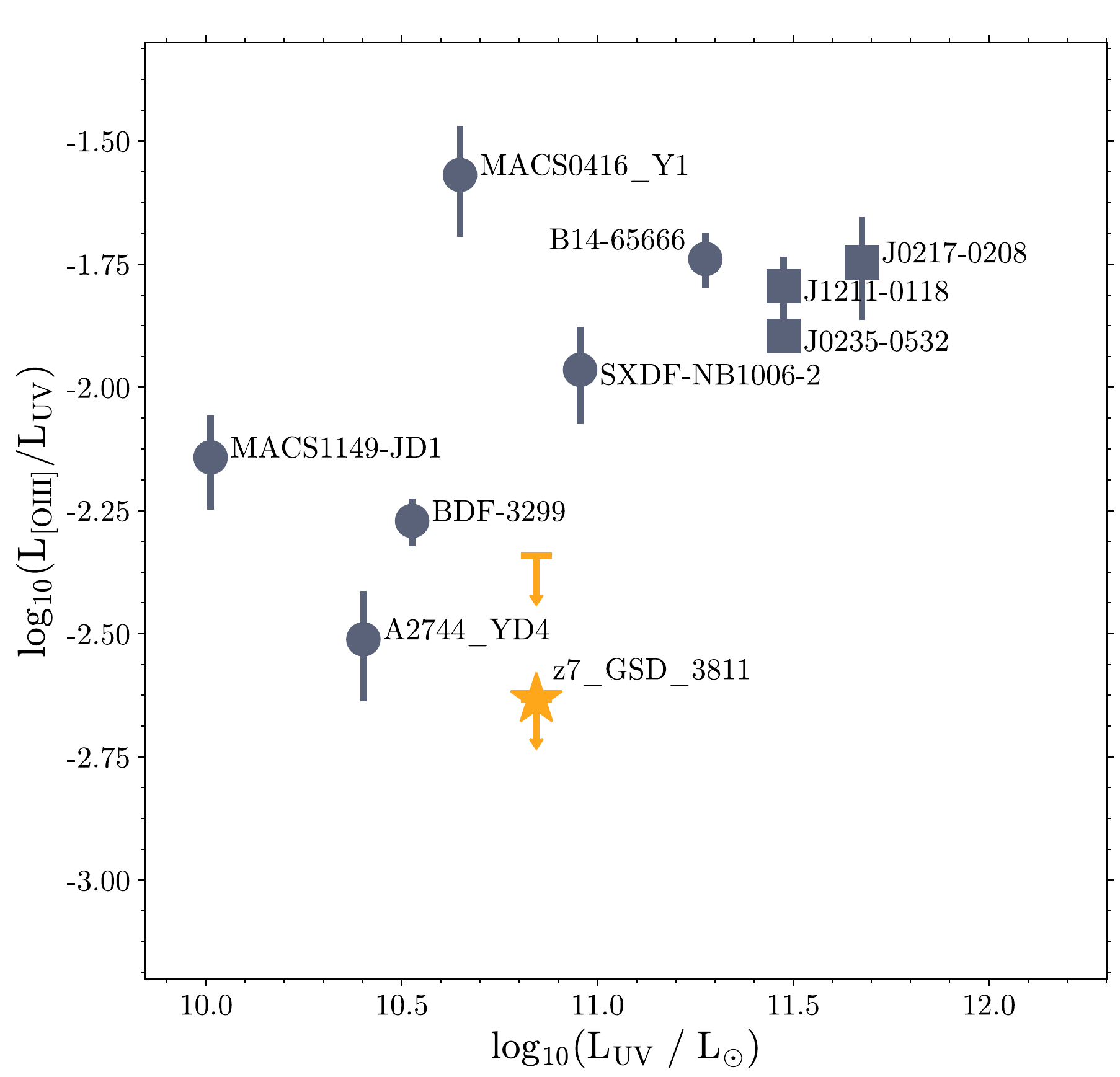}}
    \caption{The [\ion{O}{iii}] to UV luminosity ratio shown as a function of the UV luminosity. The yellow star shows the upper limit on the [\ion{O}{iii}] luminosity in z7\_GSD\_3811 assuming a line-width of $100 \ \mathrm{km \ s^{-1}}$, while the dash and arrow show the corresponding upper limit assuming a line-width of $400 \ \mathrm{km \ s^{-1}}$. The grey circles show [\ion{O}{iii}]-detected objects at $z > 7$ \citep{inoue_detection_2016,laporte_dust_2017,carniarni_extended_2017,hashimoto_onset_2018,tamura_detection_2019,hashimoto_big_2019,harikane_large_2019}, while squares show the three recent detections by \citet{harikane_large_2019} at $z\sim 6$. The UV luminosity has been calculated with lensing considered, however, the errors in the lensing magnification are not included in this figure.}
    \label{fig:L_OIII_L_UV}
\end{figure}

In Fig.~\ref{fig:L_OIII_L_UV}, we show the ratio between the [\ion{O}{iii}] and UV luminosities as a function of the UV luminosity for z7\_GSD\_3811 and a number of high-redshift detections \citep[$z\gtrsim 6$; ][]{inoue_detection_2016,laporte_dust_2017,carniarni_extended_2017,hashimoto_onset_2018,tamura_detection_2019,hashimoto_big_2019,harikane_large_2019}. While there are detections of [\ion{O}{iii}] emissions from other types of objects, such as \Edit{QSOs} and submillimeter-galaxies \citep[see e.g.][]{marrone_galaxy_2018, walter_evidence_2018, novak_ALMA_2019, hashimoto_detections_2019}, we limit this discussion to `\emph{normal}' star-forming galaxies. In order to calculate the UV luminosities of the objects shown in Fig.~\ref{fig:L_OIII_L_UV}, we use published values of $M_{UV}$ where these are available and the closest available photometry in the rest of the cases. We then calculate the UV luminosity in the same way as discussed in Sect.~\ref{sec:IRX-beta}. For B14-65666 \citep{hashimoto_big_2019} and SXDF-NB1006-2 \citep{inoue_detection_2016}, we use the values from the compilation of \citet{matthee_resolved_2019}. For BDF-3299 \citep{carniarni_extended_2017}, we use the magnitude derived in Sect.~\ref{sec:CII_SFR}. In the case of J1211-0118, J0235-0532 and J0217-0208 we use ${ M_{UV} \ = -22.8}$, $-22.8$ and $-23.3$ from \citet{harikane_large_2019}, respectively. For MACS1149-JD1 \citep{hashimoto_onset_2018}, MACS0416\_Y1 \citep{tamura_detection_2019} and A2744\_YD4 \citep{laporte_dust_2017}, we use the HST F160W magnitude \citep{zheng_young_2017}, the F140W magnitude \citep{tamura_detection_2019} and the F140W magnitude from \citet{zheng_young_2014} respectively. While we consider effects of lensing magnification, the lensing errors are not included. For A2744\_YD4, MACS1149-JD1 and MACS0416\_Y1, we use lensing magnifications of 1.8 \citep{laporte_dust_2017}, 10 \citep{hashimoto_onset_2018} and 1.43 \citep{tamura_detection_2019,kawamata_precise_2016}, respectively.

Our upper limit places z7\_GSD\_3811 below the majority of the [\ion{O}{iii}]-detected objects, and the $L_{\ion{O}{iii}} / L_{\mathrm{UV}}$ ratio from our upper limit is a factor \Edit{$\sim 12 $ and $\sim 5$} times lower than the ratio observed in  MACS0416\_Y1 \citep{tamura_detection_2019} and SXDF-NB1006-2 \citep{inoue_detection_2016}, respectively. Both of which have similar UV luminosities as z7\_GSD\_3811. The object with the closest $L_{\ion{O}{iii}} / L_{\mathrm{UV}}$-ratio is A2744-YD4 \citep{laporte_dust_2017}, in which [\ion{O}{iii}] is detected with a $\sim 4\sigma$ significance. Even if we assume a 4 times wider line, the upper limit places z7\_GSD\_3811 below earlier detections of [\ion{O}{iii}] with similar UV luminosities.

\begin{figure}
    \centering
    \resizebox{\hsize}{!}{\includegraphics[width=\columnwidth]{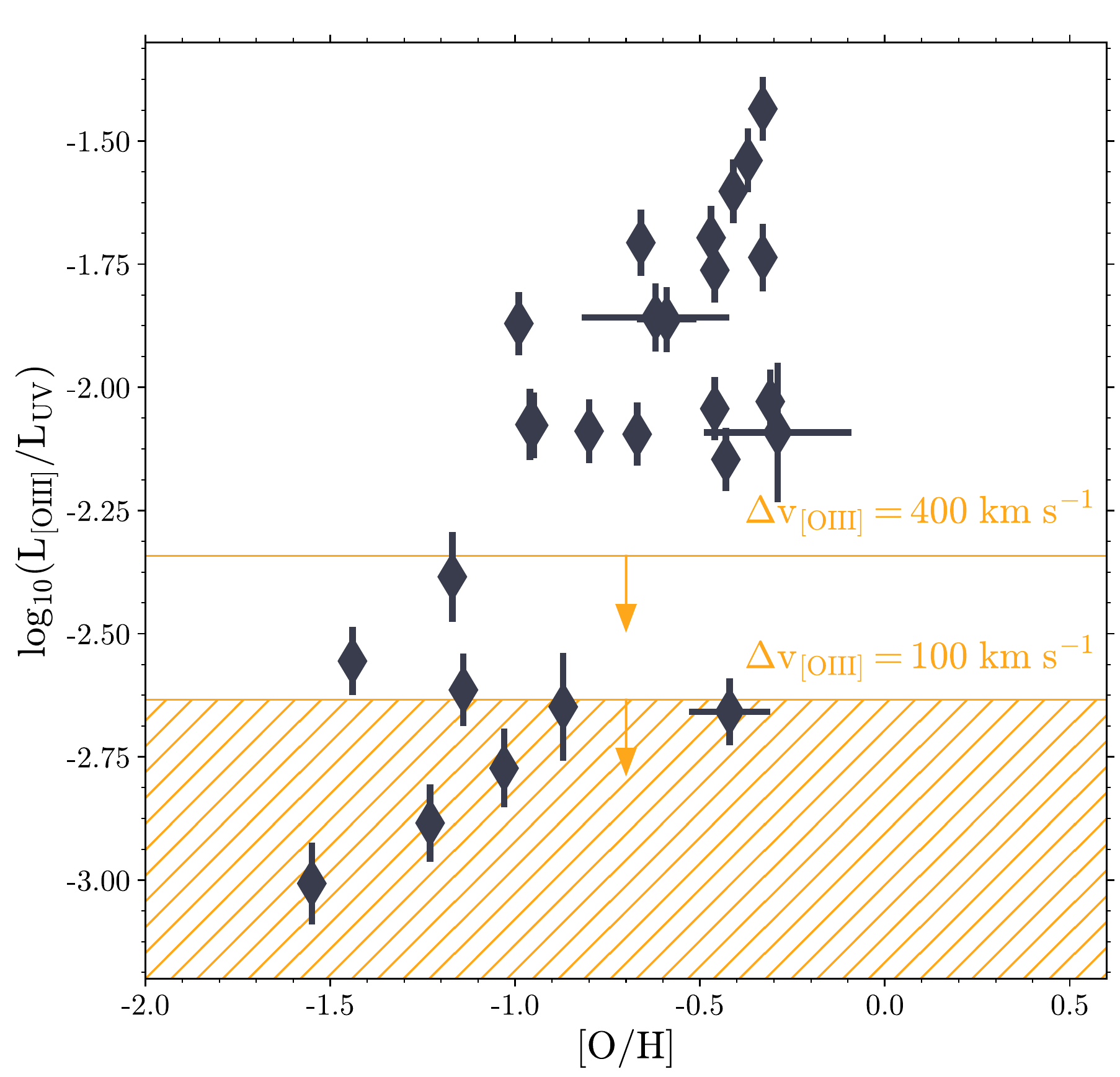}}
    \caption{The [\ion{O}{iii}] to UV luminosity ratio ($L_{\ion{O}{iii}} / L_{\mathrm{UV}}$) plotted against the oxygen abundance relative to the sun ($\mathrm{[O/H] \ = \ log_{10}}(n_O/n_H) \ - \ \mathrm{log_{10}}(n_O/n_H)_{\odot}$). The hatched area shows the upper limit on z7\_GSD\_3811 based on our $3\sigma$ upper limit on the [\ion{O}{iii}] assuming a line-width of $100 \ \mathrm{km \ s^{-1}}$, while the yellow line shows the corresponding value if we assume a 4 times wider line. The dark-gray diamonds show local dwarf galaxies from \citet{madden_overview_2013,delooze_applicability_2014,cormier_herschel_2015} compiled by \citet{inoue_detection_2016}.}
    \label{fig:L_OIII_LUV_Dwarfs}
\end{figure}

In Fig.~\ref{fig:L_OIII_LUV_Dwarfs}, we show a comparison between our upper limit on the  $L_{\ion{O}{iii}} / L_{\mathrm{UV}}$ and nearby dwarf galaxies originally from \citet{madden_overview_2013,delooze_applicability_2014,cormier_herschel_2015}, compiled by \citet{inoue_detection_2016}. In this figure, the horizontal axis shows the oxygen abundance relative to the Sun ($\mathrm{[O/H] \ = \ log_{10}}(n_O/n_H) \ - \ \mathrm{log_{10}}(n_O/n_H)_{\odot}$), where we use $12 \ + \ \mathrm{log_{10}}(n_O/n_H)_{\odot} \ = \ 8.69$ \citep{asplund_chemical_2009}. Our $3 \sigma$ upper limit places z7\_GSD\_3811 at similar $L_{\ion{O}{iii}} / L_{\mathrm{UV}}$ ratio as nearby dwarf galaxies that have an oxygen abundance of $\approx 4\%$ to $30\%$ of the solar abundance. The mean oxygen abundance of nearby dwarf galaxies with $L_{\ion{O}{iii}} / L_{\mathrm{UV}}$ lower than z7\_GSD\_3811 is $\approx 9\%$ of the solar value. This value does not change significantly in the case that we assume a four times wider line. \Edit{Note, however, that the set of nearby dwarf galaxies represents a limited sample, and contains few galaxies with high SFR and high oxygen abundance.}

\section{SED fitting}
\label{sec:SED_fitting}
In order to get an understanding of the possible stellar populations that may lead to results that are consistent with the HST, VLT and ALMA observations, we fit the SED using the \textsc{panhit} code \citep{mawatari_balmer_2019}. We use measurements from HST (F105W, F125W, F160W) and ground based infrared photometry (VLT/HAWK-I Ks) from the CANDELS catalog \citep{guo_candels_2013} combined with the ALMA upper limits on the [\ion{O}{iii}] 88 $\mu$m emission (assuming a line-width of $100 \ \mathrm{km \ s^{-1}}$) and dust continuum emission in band 6 and band 8. We do not include the upper limit on the [\ion{C}{ii}] line due to the difficulty of accurately modeling lines that receive contributions from both \ion{H}{ii} regions and PDRs \citep[e.g.][]{abel_region_2005,nagao_metallicity_2011,inoue_ALMA_2014}. For the ALMA continuum observations, we create mock-filters with a top-hat shape, with frequency width corresponding to the four SPWs in each band. The central frequency of this filter is set to be the center in the continuum image in each band. We assume a Chabrier initial mass function \citep[IMF;][]{chabrier_galactic_2003} and a single stellar component. Since z7\_GSD\_3811 remains undetected in all but four filters, we assume the simplified case of a constant SFR. The stellar models in \textsc{panhit} come from GALAXEV \citep{bruzual_stellar_2003}, nebular emission (continuum and lines) are from \citet{inoue_rest-frame_2011,inoue_updated_2014} and dust FIR emission is implemented using the empirical templates by \citet{rieke_determining_2009}. The nebular emission models from \citet{inoue_rest-frame_2011,inoue_updated_2014} are implemented in \textsc{panhit} using a scheme in which different densities are assumed for the regions where the FIR and UV/optical lines are formed. The UV/optical lines are calculated as an average over densities between $\mathrm{log_{10}}(n_{\mathrm{H}} / \mathrm{cm^{-3}}) = 0.0 \text{ -- } 2.0$, while the [\ion{O}{iii}] 88 $\mu$m line is calculated using a density of $\mathrm{log_{10}}(n_{\mathrm{H}} / \mathrm{cm^{-3}}) = 1.0$. The metallicity range used for the fitting is $Z = 0.0001$ -- $0.02$, while a \citet{calzetti_dust_2000} attenuation curve with $A_{\mathrm{V}} = 0 \text{ -- } 0.5$ with steps of $A_{\mathrm{V}} = 0.05$ was used to account for possible dust-reddening and dust emission. In the SED fitting procedure, we fix the redshift to $z=7.664$ and the escape fraction of ionizing photons to zero. This leaves us with 7 observed data-points and 4 parameters to be fitted, leading to 3 degrees of freedom. In order to account for IGM attenuation of the model spectra shortward of $\mathrm{Ly\alpha}$, \textsc{panhit} uses the model proposed in \citet{inoue_updated_2014}.
For a more in-detail description of the \textsc{panhit} SED fitting code, see \citet{mawatari_balmer_2019} and the \textsc{panhit} webpage\footnote{Panhit webpage, \url{http://www.icrr.u-tokyo.ac.jp/~mawatari/PANHIT/PANHIT.html}, maintained by Ken Mawatari.}.

\begin{figure}
    \centering
    \resizebox{\hsize}{!}{\includegraphics{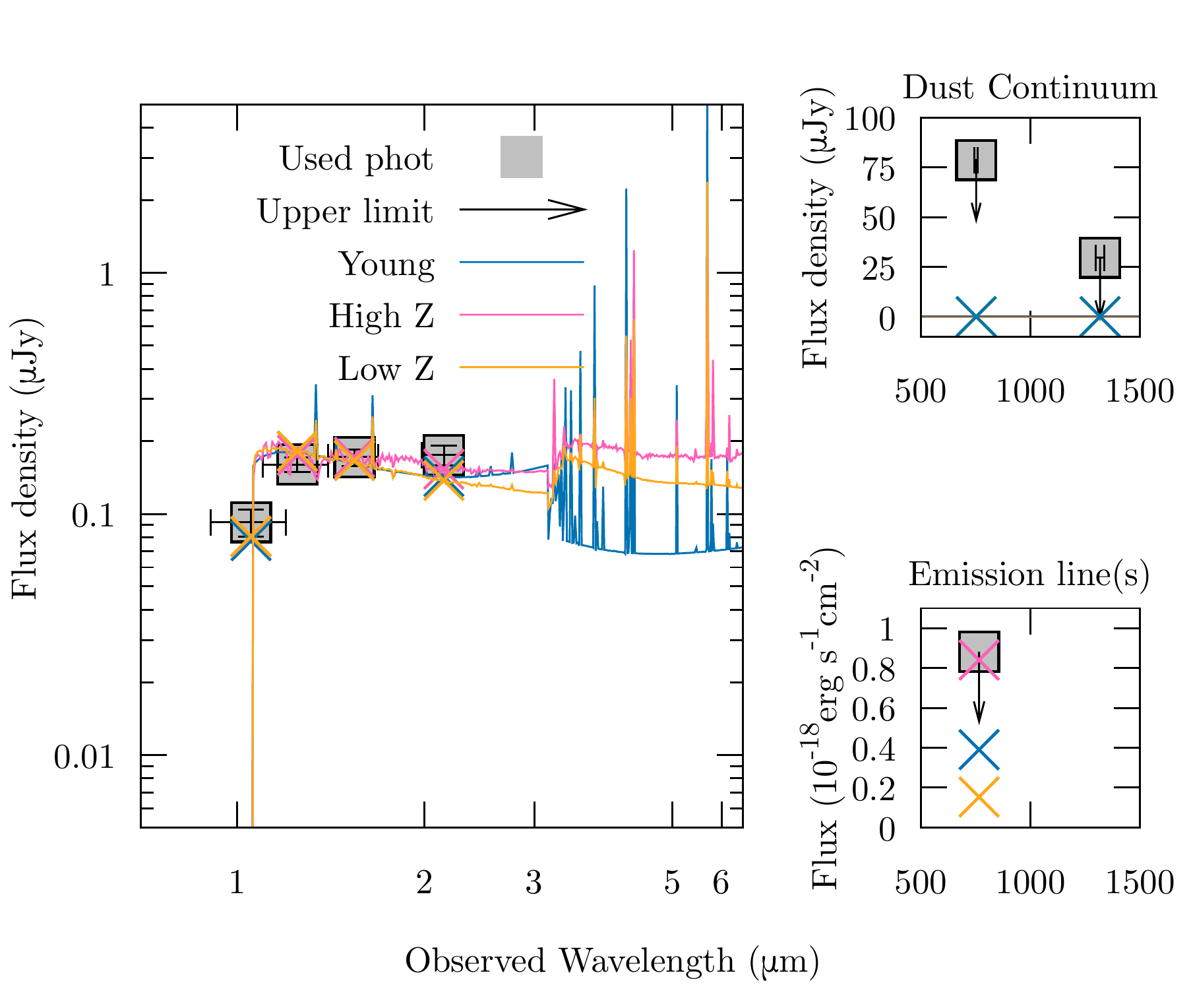}}
    \caption{\textsc{panhit} SEDs of three selected models with ages $<0.3\ \mathrm{Gyr}$ (see text). The yellow line shows an old ($0.29 \ \mathrm{Gyr}$), low-metallicity ($Z \ = \ 0.0004$) model (Low Z), the pink line shows a model with the same age but high metallicity ($Z \ = \ 0.02$ ; High Z) and the blue line shows a young, low-metallicity model (Young; Age $= \ 2 \ \mathrm{Myr}$, $Z \ = \ 0.0001$). The main panel shows the spectrum in the rest-frame UV and optical, with upper limits and measured photometry from HST and VLT/HAWK-I. The small panels show our ALMA upper limits on the dust continuum \emph{(top panel)} and the [\ion{O}{iii}] emission line \emph{(bottom panel)}. Upper limits are at the $3\sigma$ level and are indicated with arrows. Note that the dust-continuum fluxes of the three models (crosses) overlap in the top panel.}
   \label{fig:SED_panhit}
\end{figure}

Formally, the best-fitting solution (\Edit{$\chi_{\nu}^2 \  = \ 3.40$}) obtained from \textsc{panhit} gives a high age of $0.57 \ \mathrm{Gyr}$ and a low metallicity of $Z \ = \ 0.0004$. The star formation rate of the model is $\mathrm{SFR} \ = \ 7.95 \  \mathrm{M_{\odot}yr^{-1}}$, with a stellar mass of $M_{\star} \ = \ 3.05 \times 10^{9} \ \mathrm{M_{\odot}}$ and a dust-extinction of $\mathrm{A}_V \ = \ 0.00$ magnitudes. Considering the age of the Universe at $z=7.664$ is only $\sim 0.67 \ \mathrm{Gyr}$, this high age can certainly be questioned. In the context of constant SFR models, while cosmic star formation may well start at $\sim 100 \ \mathrm{Myr}$ after the Big Bang, we do not expect galaxies to maintain a constant SFR from this time until $z\sim7.7$, since simulations generally suggest increasing SFRs over time. Note that we have also performed tests with exponentially increasing SFRs, which produce fits that are not essentially different than the ones obtained from a constant SFR model. If we restrict the age of our constant SFR models to 300 Myr, as would seem more realistic given star formation histories (SFHs) of simulated galaxies \citep[e.g. the simulations by][]{shimizu_nebular_2016}, we end up with three solutions with very similar chi-square but very different properties (see Fig.~\ref{fig:SED_panhit} and table~\ref{table:SED-Fitting}). 

These three solutions are statistically indistinguishable, and their relative probabilities are sensitive to percent level variations in our [\ion{O}{iii}] upper limit. Here we find a Young solution (with an age of $2 \ \mathrm{Myr}$) with low metallicity ($Z \ = \ 0.0001$), and two older solutions (with ages of $ 0.29 \ \mathrm{Gyr}$), one with a high metallicity ($Z \ = \ 0.02$) and the other with a low metallicity ($Z \ = \ 0.0004$). Hence, the SED fitting does not uniquely favour a low metallicity solution. It should be noted, however, that while the high metallicity solution has a relatively low chi-square, the [\ion{O}{iii}] emission of the model is only \Edit{$\sim 5\%$} weaker than the observed $3\sigma$ upper limit. In the case of the young model, the relatively red UV slope obtained from the CANDELS catalog is reproduced with strong nebular emission, as seen in Fig~\ref{fig:SED_panhit}. At such a young age, all of the models with valid fits have metallicities $Z\lesssim0.0004$. Note that using an SMC dust extinction law instead of the \citet{calzetti_dust_2000} attenuation law yields no significant difference in the properties of the best-fit models because of the strong pressure toward negligible dust content set by our upper limits on the dust continuum. \Edit{We also test our results in a modified version of \textsc{panhit}, where the user can specify dust temperature and dust emissivity index. We use the same parameters as in Sect.~\ref{sec:dust_upperlimit} ($T_d = 45 \ \mathrm{K}$, $\beta_d=1.5$). While the individual chi-squares of the solutions change, we find that the favored solutions remain largely the same, with slightly higher dust contents ($\mathrm{A}_V \ = \ 0.05$--0.1).} However, if we assume a higher dust temperature \citep[$T_d = 80 \ \mathrm{K}$, as proposed to explain the recent observations of MACS0416\_Y1 by ][]{bakx_ALMA_2020}, solutions with higher dust attenuation/thermal re-emission ($\mathrm{A}_V \ = \ 0.3$--0.4) can attain acceptable chi-squares. An effect of this is that \textsc{panhit} favours low-metallicity solutions with dust-reddened continua (which, in this case, attain $\chi_{\nu}^2 \  \lesssim \ 2.7$) over high-metallicty solutions without dust.

In principle, observations of rest-frame UV/optical emission lines should provide us with a way to disentangle the different solutions suggested by \textsc{panhit}. For this purpose, we use \textsc{panhit} to estimate line luminosities for selected lines that we should be able to target with the upcoming JWST.  Lines such [\ion{O}{ii}]~$\lambda$3727, $\mathrm{H} \beta$ and [\ion{O}{iii}]~$\lambda$5007 should fall within the wavelength range of JWST/NIRSpec at the source redshift. Furthermore, H$\alpha$ should be observable with JWST/MIRI. Using the JWST exposure time calculator\footnote{\url{https://jwst.etc.stsci.edu/}} \citep{pontoppidan_pandeia:_2016}, we can estimate the time required to detect these lines with a given SNR. We set the assumed line-width to be $ 100 \ \mathrm{km \ s^{-1}}$ and use no continuum and treat the object as a point-source. The line fluxes obtained from \Edit{\textsc{panhit}} are shown in table~\ref{table:SED-Fitting}. For all three models, $\mathrm{H} \beta$ and [\ion{O}{iii}] $\lambda 5007$ are bright enough to be observed with an SNR of $\gtrsim 5$ given an exposure time of approximately one hour with the JWST/NIRSpec multi-object spectrograph using the Prism setting. In the case of [\ion{O}{ii}] $\lambda 3727$, we find that we should get an SNR of $\sim 5$ with approximately one hour of exposure for the old high-metallicity solution. For the other two solutions, the line is too weak to be detected with an SNR of $\sim 5$ within a reasonable exposure time. All three models predict $\mathrm{H\alpha}$ strengths which should be detectable with an SNR of $\sim 5$ given an exposure time of approximately six hours using MIRI medium resolution spectroscopy. \Edit{In addition, \ion{He}{ii} $\lambda 1640$ should fall within the JWST/NIRSpec wavelength range at these redshifts. This line has been suggested as a signature of very metal-poor stellar populations \citep[e.g.][]{schaerer_transition_2003,raiter_predicted_2010}, although recent results highlight that the connection between detections of \ion{He}{ii} and low metallicity is not entirely clear \citep[e.g.][]{kehrig_extended_2015,senchyna_ultraviolet_2017,berg_window_2018,kehrig_extended_2018,schaerer_x-ray_2019,senchyna_high-mass_2020}. In any case, the \ion{He}{ii} $\lambda 1640$ flux predicted for the models discussed here is not bright enough to be observed with the JWST/NIRSpec within a reasonable exposure time.}

\begin{table}
      \caption[]{Properties of the three selected SED-fitting solutions (see text) and predicted line luminosities for these. The table shows the reduced chi-square ($\chi_{\nu}^2$), the absolute metallicity ($Z$) the Age, SFR, stellar mass ($M_{\star}$), dust extinction ($A_V$) along with predicted line fluxes (in units of $\mathrm{erg \ s^{-1} \ cm^{-2}}$) and line ratios ($\left[ \ion{O}{iii} \right] \ \lambda 5007 / \left[ \ion{O}{ii} \right] \ \lambda 3727$ and $\left[ \ion{O}{iii} \right] \ \lambda 5007 / \mathrm{H} \beta$).}
         \label{table:SED-Fitting}
     $$ 
         \begin{array}{l l l l}
            \hline
            \noalign{\smallskip}
                 {} & \mathrm{High \ Z} & \mathrm{Low \ Z} & \mathrm{Young} \\
            \noalign{\smallskip}
            \hline
            \noalign{\smallskip}
            \chi_{\nu}^2 & \Editm{4.04} & \Editm{3.90} & \Editm{3.95}\\
            Z & 0.02 & 0.0004 & 0.0001 \\
            \mathrm{Age} \ (\mathrm{Gyr}) & 0.29 & 0.29 & 0.002 \\
            \mathrm{SFR} \ (\mathrm{M_{\odot}} \ \mathrm{yr^{-1}}) &  10 & 8.4 & 77 \\
            M_{\star} \ (10^9 \ \mathrm{M_{\odot}}) & 2.1 & 1.7 & 0.16 \\
            A_V \ (\mathrm{mag.}) & 0.0 & 0.0 & 0.0\\
             &  &  & \\
            \left[ \ion{O}{ii} \right] \ \lambda 3727 & 1.1 \times 10^{-18} & 1.5 \times 10^{-19} & 1.7 \times 10^{-19}\\
            \mathrm{H} \beta & 9.4 \times 10^{-19} & 1.3 \times 10^{-18} & 6.3 \times 10^{-18}\\
            \left[ \ion{O}{iii} \right] \ \lambda 5007 & 2.9 \times 10^{-18} & 1.4 \times 10^{-18} & 1.9 \times 10^{-18}\\
            \mathrm{H\alpha} & 2.8 \times 10^{-18} & 3.6 \times 10^{-18} & 1.8 \times 10^{-17}\\
            &  &  & \\
            \Editm{\left[ \ion{O}{iii} \right]/\left[ \ion{O}{ii} \right]} & \Editm{2.6} & \Editm{9.3} & \Editm{11}  \\
            \Editm{\left[ \ion{O}{iii} \right]/\mathrm{H} \beta } & \Editm{3.1} & \Editm{1.1} & \Editm{0.30} \\
            \noalign{\smallskip}
            \hline
         \end{array}
     $$ 
 \end{table}

\section{Discussion \& Summary}
\label{sec:Disc}
We have presented recent ALMA observations targeting FIR [\ion{O}{iii}] 88 $\mu \mathrm{m}$, [\ion{C}{ii}] 158 $\mu \mathrm{m}$ emission lines and dust continuum in the high-redshift galaxy z7\_GSD\_3811. The object is not detected in line nor continuum images. We therefore present upper limits of the line and dust continuum emission. In Sect.~\ref{sec:results}, we show that our UV+IR SFR and $3 \sigma$ upper limit on the [\ion{C}{ii}] luminosity would place z7\_GSD\_3811 below the \citet{delooze_applicability_2014} relations for local HII/Starburst galaxies and local low-metallicity dwarfs. 

\Edit{Similar results have been found in a number of studies targeting [\ion{C}{ii}] in high-redshift galaxies \citep[e.g.][]{Ota_ALMA_2014,Schaerer_constraints_2015,maiolino_assembly_2015,matthee_resolved_2019}, and there are several studies discussing the topic and possible explanations \citep[see e.g.][]{vallini_CII_2015, carniani_kiloparsec-scale_2018, harikane_silverrush_2018, harikane_large_2019}. However, as discussed in Sect.\ref{sec:introduction}, \citet{schaerer_ALPINE_2020} find better consistency between high-redshift and local objects and derive a [\ion{C}{ii}]-SFR relation at high redshifts which is only marginally steeper than the local relation for HII/Starburst galaxies. Our upper limit on the [\ion{C}{ii}] luminosity places z7\_GSD\_3811 below this relation, although only marginally so when we consider the uncertainties in the fitted relation by \citet{schaerer_ALPINE_2020}. As shown in Sect.~\ref{sec:CII_SFR} this does also depend on our assumption on the line-width. Formally, our upper limit could be consistent with either of the relations shown in Fig.\ref{fig:SFR_CII} given a substantial scatter. However, out of these, the slightly steeper relation derived by \citet{schaerer_ALPINE_2020} using measurements and $3\sigma$ upper from the ALPINE survey and earlier z>6 data causes the least tension with our results.}

\Edit{Other studies that have examined the [\ion{C}{ii}]-SFR relation at high redshifts have found that the [\ion{C}{ii}]/SFR ratio is anti-correlated with the rest-frame equivalent width of $\mathrm{Ly\alpha}$, albeit with a significant scatter in observed [\ion{C}{ii}]/SFR vs $\mathrm{EW(\mathrm{Ly\alpha})}$  \citep{carniani_kiloparsec-scale_2018, harikane_silverrush_2018, harikane_large_2019}. Compared to the relations in \citet{harikane_silverrush_2018, harikane_large_2019}, our [\ion{C}{ii}] upper-limit is lower than indicated by the modest $\mathrm{Ly\alpha}$ equivalent-width in z7\_GSD\_3811 \citep[$\mathrm{EW(\mathrm{Ly\alpha})}\approx 16 \ \AA$;][]{song_keck_2016}. On the other hand, \citet{schaerer_ALPINE_2020} find no such strong anti-correlation, which would seem to be in line with our findings of a modest $\mathrm{Ly\alpha}$ equivalent-width combined with weak [\ion{C}{ii}]. Of course, at these redshifts, absorption of $\mathrm{Ly\alpha}$ in the IGM is likely to have a large effect on the observed emission line. Thus, it is possible that the actual equivalent width is significantly larger.}

In our search for [\ion{O}{iii}] emission from high-redshift galaxies, z7\_GSD\_3811 represents the first non-detection, placing it at an $L_{\ion{O}{iii}}/L_{\mathrm{UV}}$ ratio which is \Edit{$\sim 5 \text{ -- } 12$} times lower than those observed in earlier high-redshift detections in galaxies with similar UV luminosities \citep{inoue_detection_2016,tamura_detection_2019}. One possible explanation for this could be that z7\_GSD\_3811 contains less metals than the other [\ion{O}{iii}] detected objects. The mean oxygen abundance of local low-metallicity dwarf galaxies that exhibit $L_{\ion{O}{iii}} / L_{\mathrm{UV}}$ lower than our upper limit is $\approx 10\%$ of the solar value \Edit{(see Fig.~\ref{fig:L_OIII_LUV_Dwarfs})}. If the results of \citet{steidel_reconciling_2016}, indicating an enhancement of the oxygen abundance at high redshift, extend to $z\sim7.7$, this could, in principle, mean that the metallicity may be even lower. \citet{steidel_reconciling_2016} argue that such an effect can be explained by a scenario in which the ISM enrichment is dominated by metals produced by core-collapse supernovae. \Edit{However, due to a lower Lyman continuum emissivity at high metallicities, the relation between [\ion{O}{iii}] 88 $\mu \mathrm{m}$ and metallicity flattens and may actually turn over around $Z\approx0.2 \ \mathrm{Z_{\odot}}$ \citep{inoue_ALMA_2014}. Thus, we emphasize that our upper limit does not exclude the possibility that the object has a higher metallicity, as the dwarf galaxy sample shown in Fig.~\ref{fig:L_OIII_LUV_Dwarfs} contains few objects at these metallicities. Although, the non-detection of dust emission suggests that a high metallicity is unlikely.}

In the case of [\ion{C}{ii}], we also expect this line to be weaker in low metallicity environments. The relation between weak [\ion{C}{ii}] vs SFR and metallicity in high-redshift galaxies is, however, still not entirely clear. For example, while \citet{harikane_large_2019} find a connection between the PDR density and CMB effects on the [\ion{C}{ii}] luminosity, they do not find a significant effect on the [\ion{C}{ii}] luminosity as a function of the metallicity in their model. Such an effect is on the other hand seen in the simulations by \citet{vallini_CII_2015}, where the authors find that lower gas metallicities could explain the deficit in [\ion{C}{ii}] vs SFR observed in some high-redshift galaxies. If we compare our upper limit to the [\ion{C}{ii}] luminosity and our UV SFR to the [\ion{C}{ii}]-SFR relation by \citet{vallini_CII_2015}, we find that the values we obtain are consistent with a metallicity of $Z\lesssim 0.08 \ \mathrm{Z_{\odot}}$ for a line-width of $\Delta v = 100 \ \mathrm{km \ s^{-1}}$. In the case that we assume a four times wider line, this upper limit is relaxed to $Z\lesssim 0.12 \ \mathrm{Z_{\odot}}$. Since the simulations used to calibrate the relationship in \citet{vallini_CII_2015} are using a \citet{salpeter_luminosity_1955} IMF, we convert our UV SFR to this IMF in order to be consistent. 

As shown in Sect.~\ref{sec:SED_fitting}, while low metallicity SED-fitting solutions are consistent with our observations, we find that these are not exclusive. Indeed, assuming a `modest' SFR and relatively high age of ($0.29 \ \mathrm{Gyr}$), solutions with approximately solar metallicity, in the case that the [\ion{O}{iii}] line luminosity lies just below our ALMA upper limit, are also derived from the SED-fitting with similar probabilities. The low $L_{\ion{O}{iii}}/L_{\mathrm{UV}}$ ratio compared to high-redshift objects with similar UV luminosities \citep[SXDF-NB1006-2 and MACS0416\_Y1; ][]{inoue_detection_2016,tamura_detection_2019} could thus also be explained by a difference in the SFHs of these objects. 

When we take the SED fitting results in conjunction with our [\ion{C}{ii}]-SFR and $L_{\ion{O}{iii}} / L_{\mathrm{UV}}$ compared to observations of local galaxies, this suggests that a likely explanation to the weak [\ion{C}{ii}], [\ion{O}{iii}] and dust continuum emission is that z7\_GSD\_3811 has a low metallicity. \Edit{In addition, objects with stellar masses around $10^9 \ \mathrm{M_{\odot}}$ and metallicities of $0.5 \text{ -- } 1 \ \mathrm{Z_{\odot}}$ should be rare at these redshifts according to the simulation of \citet{shimizu_nebular_2016}.} As discussed in Sect.~\ref{sec:IRX-beta}, the CANDELS catalog fluxes indicate a slightly red UV slope. Thus, the non-detection in the dust continuum `forces' \textsc{panhit} to increase the age of the object in order to match the UV slope, due to the strong pressure toward negligible dust content set by our upper limits on the dust continuum. With the limited number of observations in the rest-frame optical and UV, it is difficult to accurately constrain the SFH and age (and therefore also the SFR) of the object. A result of this is that we find a solution that exhibits a significantly lower age ($2 \ \mathrm{Myr}$) that also fits the observations. This solution is similar to the ones suggested for SXDF-NB1006-2 and MACS0416\_Y1 \citep{inoue_detection_2016,tamura_detection_2019}. In this case, the flat/red UV slope is explained by nebular continuum emission. This effect is also expected in the case of very metal-poor starburst galaxies, including the extreme case of a population III -dominated object \citep{raiter_predicted_2010}. If the SFR actually \Edit{is} as high as predicted by this solution ($\mathrm{SFR} \ = \ 77 \  \mathrm{M_{\odot}yr^{-1}}$), this could indicate that the object contains very little metals. As we discuss in Sect.~\ref{sec:SED_fitting}, JWST will likely be able to detect $\mathrm{H} \beta$ and $\mathrm{H\alpha}$ in z7\_GSD\_3811, which should give us a much better handle on the SFR and possibly also on the metallicity given our ALMA upper limit on the [\ion{O}{iii}] line. 

There are, of course also other mechanisms that could lead to weak [\ion{O}{iii}] emission, such as high escape fractions of ionizing photons from \ion{H}{ii} regions. However, with the limited number of observational constraints, it is difficult to meaningfully constrain the escape fraction. Additionally, since the [\ion{O}{iii}] and [\ion{C}{ii}] lines are sensitive to the density in the environment where they are formed \citep[e.g.][]{simpson_infrared_1975}, increased ISM densities could lead to both weak [\ion{O}{iii}] and [\ion{C}{ii}] emission \citep{harikane_large_2019}. Comparing our [\ion{O}{iii}] upper limit and UV SFR to the cloudy modeling by \citet{harikane_large_2019}, we find that a low metallicity ($Z \lesssim 0.2 \ \mathrm{Z_{\odot}}$), a high density ($\mathrm{log_{10}}(n_{\mathrm{H}} / \mathrm{cm^{-3}}) \gtrsim 3.0$), or a low ionization parameter ($\mathrm{log_{10}}(U_{\mathrm{ion}}) \lesssim -3.0$) is required to explain the low [\ion{O}{iii}] luminosity, \Edit{which is in line with our earlier arguments}. If it is the case that the density is high in z7\_GSD\_3811, a possible way to examine this would be to target the [\ion{O}{iii}] 52 $\mu \mathrm{m}$ emission line, since this line is expected to be stronger than the 88 $\mu \mathrm{m}$ line at high densities \citep{simpson_infrared_1975, Pereira-Santaella_Far-infrared_2017}. In addition, our FIR dust continuum observations indicate that z7\_GSD\_3811 likely contains relatively little dust. This is also consistent with the results from our SED fitting, where a majority of the best-fit models contain no dust. These results do, as shown in Sect.~\ref{sec:SED_fitting}, of course, depend on the assumed dust temperature at high redshifts. If dust temperatures as high as 80 K \citep[as recently suggested by][]{bakx_ALMA_2020} are common at high redshifts, z7\_GSD\_3811 could actually be subject to higher dust extinction.

As discussed in Sect.~\ref{sec:ALMA_b6_b8_obs}, we estimate a relative uncertainty on the flux density given the variations in spectral index observed for our flux calibrators. While there is already an uncertainty on the [\ion{O}{iii}] luminosity coming from our line-width assumption, adding 30\% to our $3\sigma$, $\Delta v = 100 \ \mathrm{km \ s^{-1}}$ upper limit would lead to a $L_{\ion{O}{iii}}/L_{\mathrm{UV}}$ ratio of $\mathrm{log_{10}(L_{\ion{O}{iii}}/L_{\mathrm{UV}})} \ \approx \  -2.5$. This would put our upper limit at a similar $L_{\ion{O}{iii}}/L_{\mathrm{UV}}$ ratio as A2744\_YD4, but still significantly lower than that observed in MACS0416\_Y1 and SXDF-NB1006-2, which have similar UV luminosities. Similarly, the position of z7\_GSD\_3811 relative to local dwarf galaxies would not change significantly, and would still be consistent with an oxygen abundance of $\lesssim 10\%$ of the solar value. For our band 6 data, the possible additional $10\%$ uncertainty from the flux calibration does not have a significant effect on the results presented here. A strategy currently tested by the ALMA Observatory, where the flux calibration is based on antenna efficiencies has the potential to be more accurate than flux calibration using QSO, especially at higher frequencies, (see the ALMA Technical Handbook:  \url{https://almascience.eso.org/documents-and-tools/cycle7/alma-technical-handbook}) and thus may help to reduce the contribution of flux uncertainty in future analysis.

In summary, our ALMA results show that z7\_GSD\_3811 is very weak in [\ion{O}{iii}], [\ion{C}{ii}] and FIR dust emission. While the nature of the object remains uncertain, these results may suggest that z7\_GSD\_3811 has a low metallicity ($Z \lesssim 0.1 \ \mathrm{Z_{\odot}}$) and little dust. This could indicate that the object is in an unevolved phase of chemical evolution and that we are revealing the earliest phase of the galaxy formation process. Furthermore, the lack of a detected [\ion{O}{iii}] emission line shows that there may be a significantly larger spread in [\ion{O}{iii}] versus UV luminosity than indicated by earlier observations. \Edit{If we assume that these objects such as z7\_GSD\_3811 are more common than currently indicated, we can expect future ALMA studies of [\ion{O}{iii}] to result in more non-detections.} As concerns z7\_GSD\_3811, future JWST observations should allow us to get a better handle on the current and previous star formation and possibly the metallicity through observations of rest-frame UV/optical continuum and emission lines. This may ultimately clarify the nature of this object and its low [\ion{O}{iii}] luminosity compared to similar high-redshift objects.

\begin{acknowledgements}
C. Binggeli would like to thank A. Gavel, A. Lavail and M. Sahlén for helpful discussions. M. C. Toribio would like to thank A. M. S. Richards, E. Fomalont and L. T. Maud for useful discussions. A. K. Inoue and K. Mawatari acknowledge the support from the Japan Society for the Promotion of Science (JSPS) KAKENHI Grant Numbers 26287034 and 17H01114. A. K. Inoue and T. Hashimoto acknowledge support from the National Astronomical Observatory of Japan (NAOJ) ALMA Grant 2016-01A. T. Okamoto acknowledges the support from JSPS KAKENHI Grant Number 19H01931. Y. Tamura acknowledges support from NAOJ ALMA Scientific Research Grant Number 2018-09B and JSPS KAKENHI Grant Number 17H06130. This publication has received funding from the European Union’s Horizon 2020 research and innovation programme under grant agreement No 730562 [RadioNet]. This publication makes use of the following ALMA data: ADS/JAO.ALMA\#2017.1.00190.S, ADS/JAO.ALMA\#2015.1.00821.S. ALMA is a partnership of ESO (representing its member states), NSF (USA) and NINS (Japan), together with NRC (Canada), NSC and ASIAA (Taiwan), and KASI (Republic of Korea), in cooperation with the Republic of Chile. The Joint ALMA Observatory is operated by ESO, AUI/NRAO and NAOJ. The authors acknowledges support from the Nordic ALMA Regional Centre (ARC) node based at Onsala Space Observatory. The Nordic ARC node is funded through Swedish Research Council grant No 2017-00648. This research has made use of the VizieR catalogue access tool, CDS, Strasbourg, France \citep[DOI: 10.26093/cds/vizier; ][]{ochsenbein_vizier_2000}, Astropy, \footnote{http://www.astropy.org} a community-developed core Python package for Astronomy \citep{astropy_2013, astropy_2018}, APLpy, an open-source plotting package for Python \citep{aplpy_2012} and NASA’s Astrophysics Data System (ADS).
\end{acknowledgements}

\bibliographystyle{aa}
\bibliography{bibliography.bib}

\end{document}